\documentclass{aa}  

\usepackage{graphicx}
\usepackage{txfonts}
\usepackage{natbib}
\usepackage{url}
\usepackage{xcolor}
\bibpunct{(}{)}{;}{a}{}{,}%
\begin{document}

   \title{Synthetic catalog of black holes in the Milky Way}

   \author{A. Olejak
          \inst{1} \inst{2} \inst{3}
          K. Belczynski\inst{2}
          \and
          T. Bulik \inst{1}
          \and
          M. Sobolewska  \inst{4}
          }

   \institute{Astronomical Observatory, Warsaw University Ujazdowskie 4, 00-478 Warsaw, Poland\\
              \email{a.olejak@student.uw.edu.pl}
         \and
             Center for Theoretical Physics, Polish Academy of Sciences, Al. Lotnikow 32/46, 02-668 Warsaw, Poland
        \and
             Nicolaus Copernicus Astronomical Center, Polish Academy of Sciences, ul. Bartycka 18, 00-716 Warsaw, Poland\\
             \email{chrisbelczynski@gmail.com }
        \and
             Harvard-Smithsonian Center for Astrophysics,  60 Garden St, Cambridge, MA 02138, USA\\
             }

   \date{Received August 23, 2019; }

% \abstract{}{}{}{}{} 
% 5 {} token are mandatory
  \abstract
    % aims heading (mandatory)
  {}
  % aims heading (mandatory)
  {We present an open-access database which includes a synthetic catalog of
black holes (BHs) in the Milky Way, divided into components: disk, bulge and halo.}
  % methods heading (mandatory)
{To calculate evolution of single and binary star we used updated population synthesis code StarTrack. We applied a new model of star formation history and chemical evolution of Galactic disk, bulge and halo synthesized from observational and theoretical data. This model can be easily employed for farther evolutionary population studies.}
 % results heading (mandatory)
{We find that at the current moment Milky Way (disk+bulge+halo) contains about $1.2 \times 10^8$ single black holes with average mass of about 14 M$_\odot$ and $9.3 \times 10^6$ BHs in binary systems with average mass of 19 M$_\odot$. 
We present basic statistical properties of BH population in three Galactic components such as distributions of BH masses, velocities or numbers of BH binary systems in different evolutionary configurations. }
{The metallicity of stellar population has a significant impact on the final BH mass due to the stellar winds. Therefore the most massive single BH in our simulation, 113 M$_\odot$, originates from a merger of BH and helium star in a low metallicity stellar environment in Galactic halo.  We constrain that only $\sim$ 0.006 \% of total Galactic halo mass (including dark matter) could be hidden in the form of stellar origin BHs which are not detectable by current observational surveys.
We calculated current Galactic double compact objects (DCOs) merger rates for two considered common envelope models which are: $\sim$ 3-81 Myr$^{-1}$ for BH-BH, $\sim$ 1-9 Myr$^{-1}$ for BH-NS and $\sim$ 14-59 Myr$^{-1}$ for NS-NS systems. We show how DCOs merger rates evolved since Milky Way formation till the current moment having the new adopted star formation model of Galaxy.
Data files are available on our website {https://bhc.syntheticuniverse.org/}.
}


   \maketitle
%
%________________________________________________________________

\section{Introduction} \label{sec:intro}

Study of Galactic black hole population is still a big challenge as BHs themselves do not emit in electromagnetic wavelenghts, except predicted theoretically Hawking radiation. However, black holes in binary systems are in some way possible to be observed due to their interaction with their companion, for example when star transfers mass on the BH  (\url{https://stellarcollapse.org/sites/default/files/table.pdf} and references within). Binary system with BH may be identified as well by measuring velocities of components in wide binary systems \citep{2019MNRAS.486.4098I}. Recently, after constructing LIGO and Virgo detectors, double compact object systems such as BH-BH, BH-NS,NS-NS could also be detected due to the emission of Gravitational Waves \citep{2016PhRvX...6d1015A,2016PhRvL.116x1103A,2016PhRvL.116f1102A,2017PhRvL.118v1101A,2017ApJ...851L..35A,2017PhRvL.119n1101A,2018arXiv181112907T}. Unfortunately, listed methods can not be used in order to detect single black holes, which do not interact with any other massive physical objects. Therefore, the most promising way to study single black hole population seems to be gravitational lensing phenomenon \citep{2016pas..conf..121W,2016MNRAS.458.3012W}. Another possible method which could allow for detection of both isolated single and binary BHs is studying X-rays emission caused by accretion from dense interstellar medium (ISM) \citep{2018MNRAS.477..791T}. However so far there was no BH detection using this method.

  On the other hand, having an access to cosmological, population synthesis simulations and theoretical stellar evolution models one may try to predict the number of Galactic black holes in different configurations. There have been several attempts of such studies in the past. For example a total number of Milky Way black holes was estimated by \cite{1983bhwd.book.....S, 1992eocm.rept...29V,1994ApJ...423..659B,1996ApJ...457..834T}  at the level of $10^8-10^9$.  Population of Galactic BH-BH binaries has been studied by using cosmological simulations by \cite{2018MNRAS.480.2704L}. Total number of BH-BH binaries was estimated at $\sim 1.2 \times 10^6$ with the average mass 28M$_{\odot}$ per system. Massive star and double compact objects binaries were studied using galactic evolutionary code \citep{VANBEVEREN201050,2014A&A...564A.134M}. Also recently \cite{2019arXiv190711431W} predicted the number of Galactic BHs which formed from binary star systems assuming one Galactic component (disk) and constant star formation rates in Galaxy. 

At the moment, nearly twenty Galactic stellar black holes in binaries are observed \citep{2007IAUS..238....3C,2014SSRv..183..223C}, (\url{https://stellarcollapse.org/sites/default/files/table.pdf} and references within). Majority of them are in X-ray binary systems, where BH draws matter from its companion via an accretion disc. The average mass of those BHs is about 7.5 M$_\odot$.
 Recently there have been discoveries of three black hole binary systems, based on radial velocity measurements \citep{2018ApJ...856..158K,2019Sci...366..637T,Liu_2019}.
However, one need to consider that such BH binary systems are possibly not representative statistical probe as they are only a small fraction of whole Galactic BH population. Most of Galactic BHs are hard to detect as they do not interact with companion and could have had much different evolutionary history. Therefore distributions of BH properties such as masses or velocities are possibly much different than shown only by observed binaries. 

Milky Way can be divided into several main components: bulge, disk (thin and thick disk) and halo. Those parts are different in their structure, stellar properties (such as chemical composition), formation history or dynamics. For example in Galactic disk one may find young stars with high metal content, while Galactic halo is rather dominated by old and metal poor stars. In simulations we need to take into account metallicity and age distributions of Galactic stellar populations based on the recent literature as it strongly influence the course of the evolution of single and binary star systems.

In this article we only consider evolution of isolated single and binary stars in Milky Way. In particular we do not consider dynamical interactions between stars in field populations. Although such interactions are rare they may lead to some interesting results (e.g. \cite{2017MNRAS.469.3088K}).
We also do not consider triple or higher multiplicity stellar systems. Evolution of such systems may lead to formation of some exotic configurations, and may also possibly enhance BH formation (e.g., \cite{Eggleton_2001,2017ApJ...841...77A,2018arXiv180506458A} ), but typically fraction of stars in higher multiplicity systems is not too large \citep{2013ARA&A..51..269D,2010ApJS..190....1R}.
Finally, we also do not consider here Galactic globular clusters (GCs). As shown by cosmological simulations \citep{2005ApJ...623..650K} or by measuring the mass to light ratio M/L \citep{2009A&A...500..785K}, the total mass of Galactic GCs is only about $\sim$ 0.005 - 0.01 \% of the Milky Way stellar mass.
 
 Our main motivation was to create an open-access database which contain basic statistical properties of BHs in the Milky Way. Such a catalog may be useful for observers as major part of Galactic black hole population (as it is shown by previous and our current results) is so far undetected. In our catalog we list most common BH configurations, BH numbers, masses and velocities, and their place of origin. Such information may help to guide current and future electromagnetic (e.g., Gaia) and gravitational-wave (e.g., LISA) missions to detect large number of BHs in Milky Way. For double compact objects (DCOs: NS-NS/BH-NS/BH-BH) we additionally list their current merger rates in Milky Way (or similar galaxies) as they may be of some importance for LIGO/Virgo missions.

%                                     Two column figure (place early!)
%______________________________________________ Gamma_1 (lg rho, lg e)

%__________________________________________________ One column table

\section{Method} \label{sec:method}

To calculate evolutionary scenarios of star systems we used updated population synthesis method implemented in StarTrack code. Currently the code allows to simulate single star as well as a binary system evolution for a wide range of initial conditions and physical parameters. Physics formulas and methods implemented in StarTrack code have been expanded and updated over the years \citep{2002ApJ...572..407B,2008ApJS..174..223B,2017arXiv170607053B}.\\

\subsection{Initial conditions} \label{sec:initaial}

For initial mass of the single star we adopted 3-broken power-law initial mass function (IMF) from \cite{1993MNRAS.262..545K} however we calibrated power-law exponent $\alpha_3$ to match observations for massive stars as proposed by \cite{2002Sci...295...82K} : \\
 $\alpha_1= -1.3$ for  M $\in [0.08,0.5]$ M$_{\odot} $ \\
 $\alpha_2 = -2.2$ for  M $\in [0.5,1.0]$ M$_{\odot} $ \\
 $\alpha_3 = -2.3$ for  M  $\in [1.0,150.0]$ M$_{\odot} $ \\
 We apply calibrated IMF for both single stars and primaries (more massive) stars in binary systems. Mass of the secondary (less massive) star of binary system ($\mbox{M}_2$) is the mass of the more massive star ($\mbox{M}_1$) from IMF multiplied by mass ratio factor q from uniform distribution in the range $q \in [0.08/\mbox{M}_1,1]$.  

 To generate initial semi-major axis of binary system we used third Kepler law (semi-major axis-period relation). We adopted power low distribution in log(P) in range log(P[days]) $\in [0.15,5.5]$ with exponent $\alpha_P = -0.55$ and power-law initial eccentricity distribution with exponent $\alpha_e = -0.42$ in a range [0,0.9]. Such initial distribution of orbital parameters is taken from \cite{Sana1207}, however we adopted extrapolation of the orbital period distribution proposed by \cite{2015ApJ...814...58D}. These authors extended the possible period range to log(P[days]) = 5.5 in order to match current observational data for massive binary star systems, see the discussion in \cite{2014ApJS..215...15S} and in \cite{2015ApJ...814...58D}. 

In our simulation we generated single and binary star systems with different metal contents and ages. For each of 18 stellar populations of Galactic components (see Sec. \ref{sec:sfr}) with given age and metallicity we evolved 2.5 $\times$ $10^6$ binary systems (primary component in mass range 5–150 M$_\odot$ and secondary in mass range 0.08–150 M$_\odot$) and 5.0 $\times$ $10^6$ single stars (in mass range 5.0–150 M$_\odot$). The total simulated mass of systems generated for each stellar population in the whole IMF mass range 0.08-150 M$_\odot$ is 3.1$ \times 10^8$ M$_\odot$ for binaries and 7.6$ \times 10^8$ M$_\odot$ for single stars. We scaled linearly number of generated systems to fit the real mass fractions and binarity of stellar populations in Galactic components. 

\subsection{Physics model} \label{sec:physicamodell}
In our simulation we apply standard updated StarTrack physical model. We use rapid supernovae (SN) model with explosions driven by instabilities with a rapid growth time of the order of 10–20 ms \citep{Fryer1204}. The model include weak pulsation pair-instability supernovae (PPSN) and pair-instability supernovae (PSN) \citep{2017ApJ...836..244W,2016A&A...594A..97B}.
We applied moderate weak PPSN model with only up to 50\% mass loss by \cite{2019arXiv190111136L} and approximate the formula for final post PPSN remnant mass (M$_{f}$) as a function of He core mass (M$_{He}$) in the following way: 

\[ \mbox{M}_{f} =
  \begin{cases}
    0.83 \mbox{M}_{He} + 6.0 \mbox{M}_\odot \quad \quad M_{He} \in [40.0,60.0] \mbox{M}_\odot\\
    55.6 \mbox{M}_\odot \quad \quad M_{He} \in [60.0,62.5] M_\odot \\
    -14.3 \mbox{M}_{He} + 938.1 M_\odot \quad \quad M_{He} \in [62.5,65.0] M_\odot
  \end{cases}
\]

To obtain remnant natal kicks after supernovae explosion we used Maxwellian velocity distribution with $\sigma$= 265 km/s \citep{2006yCat..73600974H}. BH velocities from Maxwellian distribution are multiplied by a fallback factor f$_{fb} \in (0,1)$ which is inversely proportional to the fall-back of material after a SN explosion. As a result we obtain both high and low natal kick velocities for BHs. \\
In simulations we adopted following mass transfer settings: 50\% non-conservative RLOF and 5\% Bondi-Hoyle rate accretion onto NS/BH in common envelope phase (CE).
In this work we do not consider effects of rotation on stellar evolution. \\
In our simulation we assumed star binary fraction of 50\% so 2/3 of stars in all Galactic components are formed in binary systems and 1/3 of them are single stars, as indicated by observational data \citep{2014ApJ...788L..37G, 2014ApJS..213...34K}. Chosen value may be considered as a conservative lower limit on massive stellar binarity \citep{2014ApJS..215...15S}.
We assumed that solar metallicity is Z$_\odot$=0.014 \citep{2009ARA&A..47..481A}.

\subsection{Hertzsprung gap stars in common envelope} \label{sec:models}

We calculated two CE models (marked as A and B) which represent different scenarios for binary system with Hertzsprung gap star (HG) as a donor in the CE phase. In model B we assume that all of such binary systems merge during CE and possibly create a single BH. In model A we let such a system to survive CE phase on the energy balance calculations \citep{1984ApJ...277..355W}. Currently it is not well known which scenario operates (\cite{Ivanova0402}). 

\subsection{Mergers} \label{sec:mergers}

 \begin{table}[h!]
      \caption[]{Abbreviations of object types.} \label{tab:abb}
     $$ 
         \begin{array}{p{0.5\linewidth}l}
            \hline
            \noalign{\smallskip}
             \mbox{Object type} & \mbox{Abbreviation} \\
            \noalign{\smallskip}
            \hline
            \noalign{\smallskip}
            black hole & \mbox{BH}  \\
            neutron star & \mbox{NS}  \\
            white dwarf & \mbox{WD} \\
            main sequence & \mbox{MS}  \\
            helium star & \mbox{He} \\
            Hertzsprung gap star & \mbox{HG}   \\
            asymptotic giant branch star & \mbox{AGB}  \\
            core helium burning star & \mbox{CHeB} \\
            first giant branch star & \mbox{RGB}\\
            merger with HG star in CE & \mbox{CEHG} \\
            compact object (WD, NS, BH) & \mbox{CO} \\
            double compact object & \mbox{DCO}\\
            zero main sequence & \mbox{ZAMS} \\
            binary companion  & \mbox{Comp} \\
            stars of types: Hertzprung gap star, first giant branch star, core helium burning star, early asymptotic giant branch star ot thermally pulsing asymptotic giant branch star)  & \mbox{G} \\
            \hline
         \end{array}
     $$ 
   \end{table}

A significant part of Galactic black holes could have formed in mergers of binary systems \citep{2014MNRAS.443.1319K}. The final mass of BH formed in stellar mergers depends on components masses and evolutionary types of merging objects. The amount of mass ejected from the system after its coalescence is most likely lower for dense, compact objects than for radially extended giant stars. However, stellar collisions are not well studied yet and the final products of different merger types are not fully understood. Observations as well as simulations of stellar coalescence are made only for a limited number of object types and usually low-mass stars which are not BH progenitors \citep{Lombardi_Jr__2002,2016A&A...592A.134T} or refers to dynamical collisions in stellar clusters \citep{Glebbeek_2013}. Literature, however, indicates rather low mass loss during stellar mergers which is up to $\sim 10$ \% of the system mass {\citep{Lombardi_Jr__2002,2006ApJ...640..441L, Glebbeek_2013}}. Additionally, the further evolution of the coalescence products is not well understood.   
Therefore, we adopt a simple model to estimate evolutionary types and masses of merger products.
To specify a type and structure of objects originated from different merger types we used Table \ref{tab:merger_types} (based on Tab. 2 from \cite{Hurley_2002}). The exceptions from the 
types indicated by the table are two cases: first is when the merger product is defined as 13 (NS), we classify whether a given object is a NS or a BH based on its mass. Current observations indicate that NS mass may be as high as 2.27$ ^{+0.17}_{-0.15} M_{\odot}$ \citep{2018cosp...42E2021L}. Theoretical models result in wide range of maximum NS mass: $M_{\rm NS,max}=$ 2.2 - 2.9 M$_{\odot}$ \citep{1996ApJ...470L..61K}. In our simulations we adopt maximum NS mass of M$_{\rm NS,max}=2.5$M$_\odot$. \\
The second exception are the mergers with WDs which may lead to supernovae Ia explosion leaving no remnant behind. Our criteria for supernovae Ia explosion is in accordance with the procedure in \cite{2008ApJS..174..223B}.
\color{black} \\
In order to estimate the total mass of object created after coalescence we divide stars into four main categories: main sequence stars (MS), giants (G), helium stars (He) and compact objects (CO). All our abbreviations of object types are given in Table \ref{tab:abb}. 
The total mass of the object formed in a merger ($M_{tmp}$) is then calculated according to the procedure below:

\begin{itemize}
  \item MS-MS \\
  We take a sum of the more massive star and the fraction of mass ($f_{MS} = 0.8$) of less massive star. \\
  if ($M_{MS_1} > M_{MS_2}$): $M_{tmp} = M_{MS_1} + f_{MS} \cdot M_{MS_2}$ \\
  \item{He-He} \\
  We take a sum of the more massive star and the fraction of mass $(f_{He} = 0.8)$ of less massive star. \\
  if ($M_{He_1} > M_{He_2}$): $M_{tmp} = M_{He_1} + f_{He} \cdot M_{He_2}$ \\
  \item{G-G} \\
  We take a sum of the more massive star and the fraction of mass $(f_{G} = 0.5)$ of less massive star. \\
  if ($M_{G_1} > M_{G_2}$): $M_{tmp} = M_{G_1} + f_{G} \cdot M_{G_2}$ \\
  \item{CO-CO} \\
  We take a sum of the compact objects masses. \\
  $M_{tmp} = M_{CO_1} +  M_{CO_2}$ \\
  \item {MS-He} \\ We take a sum of mass of the He star and the fraction of mass ($f_{MS} = 0.8$) of MS star.  \\
  $M_{tmp} = M_{He} + f_{MS} \cdot M_{MS}$ \\
  \item {MS-G} \\ We take a sum of mass of the main sequence star and the fraction of mass ($f_{G} = 0.5$) of giant star. \\
  $M_{tmp} = M_{MS} + f_{G} \cdot M_{G}$ \\
  \item{MS-CO} \\
  We take a sum of the CO mass and the fraction ($f_{MS} = 0.8$) of MS star. \\
  $M_{tmp} = M_{CO} + f_{MS} \cdot M_{MS}$  \\
  \item{He-G} \\
  We take a sum of the helium star mass and the fraction of mass ($f_{G} = 0.5$) of G star. \\ 
  $M_{tmp} = M_{He} + f_{G} \cdot M_{G}$  \\
  \item{He-CO} \\
  We take a sum of the CO mass and the mass fraction of mass ($f_{He} = 0.8$) of He star.
  $M_{tmp} = M_{CO} + f_{He} \cdot M_{He}$  \\
  \item{CO-G} \\
  We take a sum of the CO mass and the fraction of mass ($f_{G} = 0.5$) of G star. 
  $M_{tmp} = M_{CO} + f_{G} \cdot M_{G}$  \\
\end{itemize}
Next, after determining the type and mass of the object formed in merger by using the scheme above, if the object is not compact (WD, NS or BH) yet we evolve it as a single star of given type and mass with the StarTrack code. We set formed star at the beginning of its given evolutionary step. Then the stellar structure (the core and the envelope mass) is calculated with the procedures given by \cite{2000MNRAS.315..543H,Hurley_2002, 2008ApJS..174..223B}. 
\color{black}

\subsection{Star formation rate of the Milky Way} \label{sec:sfr}

\begin{figure}[hbt!] 
 \centering
 \includegraphics[width=95mm]{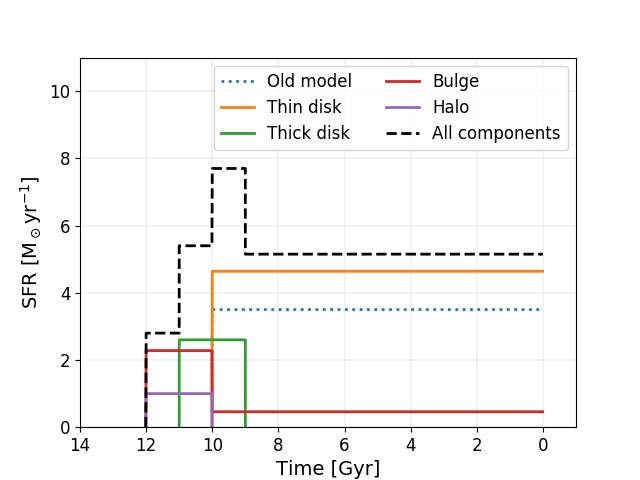}
 \caption{Star formation rate of all Galactic components as a function of lookback time. Current time is 0. The old SFR model (blue dotted line) has been plotted for comparison. }
 \label{fig:all}
\end{figure}

We created a new model of star formation rates (SFR) and metallicity distribution of stellar populations in the Milky Way.
Previously used model was not realistic as it typically considered only one Galactic component (disk) with constant SFR and same metallicity of all stars, usually equal Z=0.02 (e.g. in \cite{2010CQGra..27q3001A,2018arXiv181210065B}). Those assumptions are inconsistent with current knowledge about the Milky Way.
SFR is important for many reasons. For example the number of Galactic compact objects (BHs, NSs, WDs) scales linearly with assumed stellar mass. Formation time of the considered stellar population may significantly affects double compact objects merger rates because it determines what fraction of binaries have already merged.  \\
Metallicity of stellar populations is extremely important for BHs, as their
formation and mass depend critically on stellar winds which in turn are
a sensitive function of chemical compostion of progenitor stars (e.g.
\cite{2010ApJ...714.1217B,2014A&A...564A.134M,2018A&A...619A..77K})
We have prepared a new model of star formation rates and metallicity evolution of three Galactic components: disk (thin and thick), bulge and halo, basing on information available in the recent literature. 
Our overall model for Milky Way star formation is shown in Figure~\ref{fig:all}, 
while details are described in the following sections: for bulge see 
Sec.~\ref{sec:bulgesfr}, for disk see Sec.~\ref{sec:disksfr} and for halo 
see Sec.~\ref{sec:halosfr}. \\

\subsubsection{Galactic bulge} \label{sec:bulgesfr}
\begin{figure}[hbt!] 
 \centering
 \includegraphics[width=90mm]{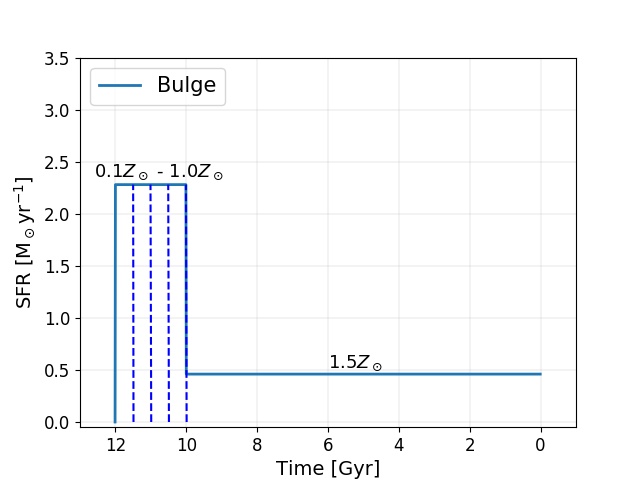}
 \caption{Star formation rate and metallicity in Galactic bulge as a function of lookback time. Current time is 0. Vertical dashed lines separate star populations with different metallicities.}
 \label{fig:bulge}
\end{figure}

Galactic bulge is the central part of Galaxy with corotation radius of $\sim 4$ kpc \citep{2001ASPC..230...21G}. In our calculation we assumed that total stellar mass of Galactic bulge is $0.91 \pm 0.07 \times 10^{10} M_\odot$ (\cite{Licquia1506}). Stellar metallicity and age distribution covers a wide range of values what is shown both by observations (e.g. \cite{2018IAUS..334...86B}) and by cosmological simulations (e.g. \cite{2011ApJ...729...16K}). To reconstruct properties of the stellar populations in bulge we used figures from \cite{2011ApJ...729...16K}. For stars age distribution we used Fig. 6, for stellar age-metallicity relation we used Fig. 8. \\ 
We approximate metallicity and age relations in the following way:\\
-half of the stars in bulge formed 10-12 Gyr ago (a formation peak, SFR=$\sim$2.3 M$_\odot$ yr$^{-1}$) and another half of stars formed from 10 Gyr ago to the current time with a constant star formation rate of $\sim 0.5$ M$_\odot$ yr$^{-1}$.\\
-star systems which are older than 10 Gyr are divided into four equal in mass groups of metallicities: 0.1 Z$_{\odot}$, 0.3Z$_{\odot}$, 0.6Z$_{\odot}$, 1.0Z$_{\odot}$ \\
-stellar population younger than 10 Gyr ago has high metal content, equal 1.5Z$_{\odot}$\\
Our model of star formation and metallicity for bulge is shown in the Figure \ref{fig:bulge}.

\subsubsection{Galactic disk} \label{sec:disksfr}

In our calculations we use the total mass of the Galactic disk $5.17 \pm 1.11 \times 10^{10}$ M$_\odot$  estimated by \cite{Licquia1506}. We assume that disk is divided into two main components: thick and thin disk which had different star formation history. Stellar populations in disk have different metal contents and ages. Thin disk is the dominant in mass part which contains about 90\% of all disk stars (\cite{2006A&A...459..783C}). Therefore, in our model  90\% of total disk stellar mass is contained in thin disk and the remaining 10\% of mass is in thick disk.

Thick disk formed as a first and its average age has been estimated for 9.6 Gyr $\pm$ 0.3 by \cite{2005A&A...438..139S}. The contribution of stars younger than 9 Gyr in thick disk is minimal as shown by \cite{2006A&A...459..783C}. In our model stellar population in thick disk formed with SFR = $\sim 2.5$ M$_\odot$ during a period from 11 to 9 Gyr ago and had a metal content equal $0.25 Z_\odot$, Fig.6 (\cite{2018ApJ...862..163L}). 

Star formation history of thin disk was estimated basing on the age distribution of observed stars showed in Fig. 13 in \cite{2011A&A...530A.138C} and results of \cite{2011ApJ...729...16K}. We approximate figures by ten star formation episodes, which started 10 Gyr ago and last till current moment. New episodes happen every 1 Gyr and last for $\Delta$t = 1 Gyr with constant star formation rate equal $\sim$ 5M $_\odot$ yr$^{-1}$. Based on the metallicity-age relation presented in \cite{2013A&A...560A.109H}, \cite{2015cdem.confE...3H}, in the next formation episodes metal content of stellar populations increases and changes from $0.1 Z_\odot$ to $Z_\odot$ with rate of $\Delta$Z = 0.1 Z$_{\odot}$Gyr$^{-1}$. 

\begin{figure}[hbt!] 
 \centering
 \includegraphics[width=90mm]{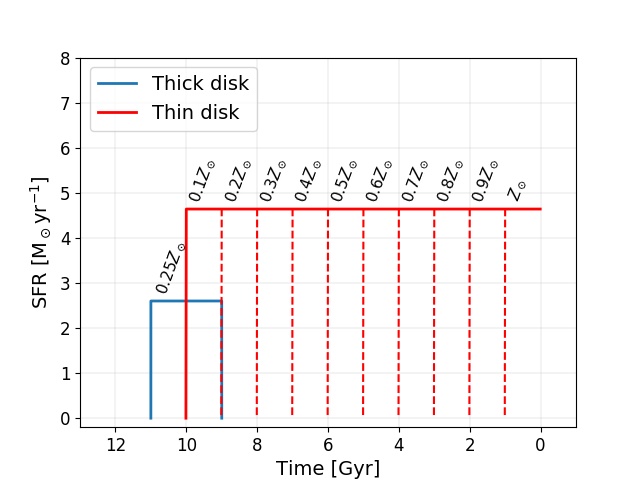}
 \caption{Star formation rate and metallicity of Galactic disk as a function of lookback time. Current time is 0. Vertical dashed lines separate star populations with different metallicities.}
 \label{figure:disk}
\end{figure}

Our SFR and metallicity distribution model of disk as a function of time is shown on Fig. \ref{figure:disk}.

\subsubsection{Galactic halo} \label{sec:halosfr}

\begin{figure}[hbt!]
 \centering
 \includegraphics[width=90mm]{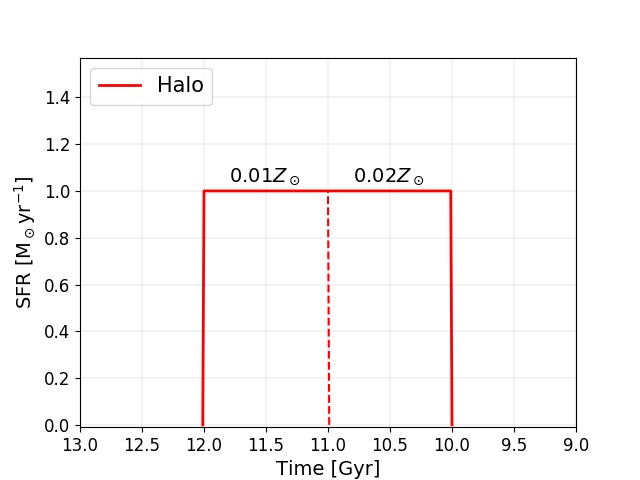} 
 \caption{Star formation rate and metallicity of Galactic halo as a function of lookback time. Current time is 0. Vertical dashed line separates star populations with different metallicities.}
 \label{fig:halo}
\end{figure}

Galactic halo is the most massive Galactic component. Total mass of halo is estimated at $\sim 10^{12} $M$_{\odot}$ \citep{2015MNRAS.453..377W,2019MNRAS.487L..72G} but most of it is dark matter, which do not reflect or emit electromagnetic radiation.
In our simulation we considered only the stellar mass in Galactic halo, which is a small fraction of the total halo mass, estimated at around $2 \times 10^9$M$_{\odot}$  \citep{2000AJ....119.2254M,2000AJ....119.2843C,2000ApJ...540..825Y,2002ApJ...578..151S,2005ApJ...635..931B}. To reconstruct age and metallicity of stars in halo we used results of cosmological simulations by \cite{2011ApJ...729...16K}. We approximate Fig. 6 to get a distribution of star ages for halo and Fig. 8 for star age-metallicity relation. \\ We simplify those relation by dividing halo into two equal in mass components which formed 11-12 Gyr and 10-11 Gyr ago with respective metallicities: 0.01Z$_{\odot}$ and 0.02Z$_{\odot}$ and star formation rate of $~\sim 0.5$ M$_\odot$ yr$^{-1}$.  \\   Model of SFR and metallicity in halo is shown on \ref{fig:halo}.

\subsection{Velocities and coordinates of Galactic black holes}

In our online database (https://bhc.syntheticuniverse.org/) we give simple estimations of Galcatic coordinates and velocities of BHs.\\
The coordinates of black holes were randomly selected using formulas for stellar mass distributions in Galactic components. For bulge and halo we used formulas given by \cite{korol2018constraining}. For bulge we draw the distances to the Galactic center from the range [0-4] kpc setting parameter rb = 1.0 kpc (the characteristic radius of the bulge), while for halo the possible distance range is [15-30] kpc. We assume that both bulge and halo are spherical. For black holes in Galactic disk we used expression from \cite{li2016modelling}. We draw distances from the range of [2-15] kpc in order to get x and y coordinates, while coordinate z is taken from the uniform distribution in range [-0.15,0.15] kpc, as we assumed average disk height 0.3 kpc \citep{Rix_2013}. Here we set parameter rb = 2.0 kpc (the  radius  of  the 
central bulge) to get continuity between disk and bulge components.

Total velocity of a Galactic BH is a sum of the motion in Milky Way gravitational potential and the velocity obtained during isolated or binary evolution. In our estimation of BH speed we considered both the rotation velocities around the Galactic center and additional BH velocities from physical processes such as SN explosions. Single and binary BHs could get a significant portion of kinetic energy after SN/core collapse formation due to asymmetric neutrinos flux release \citep{2012ApJ...749...91F,2017hsn..book.1095J}. If BH is in a binary system, the whole system changes its velocity after the first and the second NS or BH formation \citep{1996ApJ...471..352K}. If the velocity is high enough, binary system gets disrupted. Disruption of binary system may happen not only in asymmetric SN explosions. Disruption may also be a result of Blaauw kick \citep{1961BAN....15..265B} associated with symmetric mass loss which leads to change in mass ratio and orbital elements of the system. \\
 We calculated velocities of single and binary BHs after their formation in three spatial dimensions,  V$_{{S}}$=[V$_{{Sx}}$,V$_{{Sy}}$,V$_{{Sz}}$]. \\
To estimate the motion of BHs in the Milky Way potential we used approximated form of Galactic rotational curve (V$_r$), applied in other Galactic simulations (e.g. \cite{2018ApJ...868...17A}). The simplified rotational curve has a form:
\[ \mbox{V}_r =
  \begin{cases}
    220 \mbox{ km/s}  \quad \mbox{   for disk and halo}\\
    {220 \mbox{ km/s} \times \mbox{r} / \mbox{R}_b} \quad \mbox{   for bulge}
  \end{cases}
\]

In our estimates R$_b$ is the radius of bulge equal to 4 kpc \citep{2001ASPC..230...21G} and r is a distance from a BH to the Galactic center generated from the bulge mass distribution \citep{korol2018constraining} as already mentioned. \\
In order to calculate the sum of two velocities (V$_{S}$ ,V$_{r}$) we generated components of vector V$_{r}$ in three spatial dimensions for bulge and halo (spherical motion) and in two dimensions for disk (motion in plane). The total BH velocity in our results is norm of the sum of two velocities V$_{S}$ and V$_{r}$ in three dimensions.      

\section{Results} \label{sec:Results}

We present results of our simulations for Galactic components: Section \ref{sec:bulge} for bulge, Section \ref{sec:disk} for disk (thin and thick), Section \ref{sec:halo} for halo. In each of subsection one may find tables for single and binary black holes which contain basic statistical information such as estimated number of BHs in different configurations, formation channels and average masses of BHs and their companions (for binary BHs). The abbreviation of objects types are given in Table \ref{tab:abb}. In the tables there are always two values which refer to two considered evolution models A and B (see Sec. \ref{sec:models}) given in order A/(B).

In the table with single black holes we include sections: single stars, mergers and disrupted binaries which refer to BH formation channel.  Section single stars is for BHs which are remnants of massive single stars evolution. Mergers section is for single black holes created in the coalescence of binary system and is divided into several rows corresponding to types of objects that have merged. The third section is for BHs from disrupted binary systems. Note that in the merger section, in the row with BH-BH systems, the entry refers to the number of single black holes that formed in merger of two BHs while in the section for disrupted systems, the number in the row BH-BH is the total number of single black holes after system disruption (so two times the number of disrupted BH-BH binaries). \\
In the table with binary system BHs one may find two sections: one with double compact objects systems (BH-BH, BH-NS, BH-WD) and one with other binary systems in which black hole companion is unenvolved star. In table is an information about the types of BH companion objects, estimated number of BHs in given Galactic component and their average masses.\\
In Section \ref{darmatter} we constrain the amount of dark matter that could be hidden in Galactic halo in the form of stellar
origin BHs which are are not detectable by current observation
surveys . \\
In Section \ref{sec:merger_rates} we calculated how Galactic compact object merger rates have changed since Milky Way formation till current time. In Table \ref{tab:mergerates} we present current Galactic merger rates for BH-BH, BH-NS and NS-NS binary systems. \\
In Section \ref{sec:mass_distribution} we present and discuss distribution of single and binary BH mass in Galactic bulge, disk and halo. \\
In Section \ref{sec:velocity} we show and discuss velocity distribution of single and binary BHs in Galactic components. In Table \ref{tab:velocities} we give BH average speeds in bulge, disk and halo. We constrain the fraction of BHs with speed high enough to escape from Galaxy.

\begin{figure}[hbt!] 
 \centering
 \includegraphics[width=90mm]{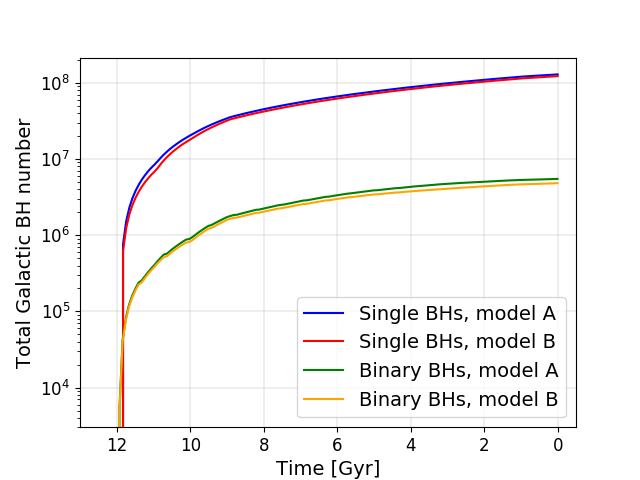}
 \caption{ Total number of formed Galactic BHs single and binary as a function of time. Current time is 0. Results for two calculated models A and B (Sec. : \ref{sec:models}).}
 \label{fig:BH_number}
\end{figure}

On Figure \ref{fig:BH_number} we show how the total numbers of single BHs and BHs in binary systems in the Milky Way have been changing since the Milky Way formation till current time for two evolutionary models A and B (Sec. \ref{sec:models}).

\subsection{Galactic bulge} \label{sec:bulge}

We find that Galactic bulge hosts about $1.7 \times 10^7$ single BHs and about $1 \times 10^6$ BHs in different types of binary systems. The average mass of single BH is 11.7/(11.3) M$_{\odot}$ while average mass of a black hole in binary system is $\sim$ 15.6/(15.8) M${_\odot}$. The most massive black holes in bulge are formed in coalescence of massive binary system (BH and massive star or binary BH system) and they can reach value of $\sim$ 100 M$_{\odot}$.

Significant amount of single black holes in bulge (nearly 50\%) are remains of single massive stars evolution. Average mass of such a BH is $\sim$ 11.7 M$_\odot$. The second important origin of single BHs are binary system mergers, especially MS-MS and MS-He. Almost 30\% of bulge single black holes are formed in mergers and their remaining masses cover a wide range of values. The least massive single black holes originates from mergers of low mass binaries such as WD-MS or WD-He. Formed masses often fill the first observational mass gap between 3-5 M$_{\odot}$ \cite{2010Natur.467.1081D, 2013Sci...340..448A,Swihart_2017,2019arXiv190407789W}. The third main formation channel of single BHs is disruption of binary systems after one of BH/NS formation. Such black holes are about 20 \% of all bulge single BHs with their average mass of $\sim$ 10 M$_{\odot}$.

   \begin{table}
      \caption[]{Single black holes in bulge\label{tab:singlebulge}}
     $$ 
         \begin{array}{lrr}
            \hline
            \hline
            \noalign{\smallskip}
             \mbox{Object type} & \mbox{Number} & \bar{\mbox{M}}_{\mbox{}BH} [\mbox{M}_\odot]\\
            \noalign{\smallskip}
            \hline
            \hline
            \noalign{\smallskip}
            \multicolumn{3}{c}{\mbox{Single stars}}    \\
            \hline
            \hline
            \mbox{Single star} & 8.5 \times 10^6/(8.5 \times 10^6)  & 11.7/(11.7) \\
            \hline
            \hline
            \multicolumn{3}{c}{\mbox{Mergers}} \\
            \hline
            \hline
            \mbox{BH - BH} & 2.0 \times 10^5 /(2.1 \times 10^4) & 28.4/(24.8) \\
            \mbox{BH - NS} & 1.6 \times 10^4 /(2.5 \times 10^3) & 10.7/(12.0) \\
            \mbox{NS - NS} & 5,0 \times 10^3 /(4,9 \times 10^2) & 2.8 /(2.6)  \\
            \mbox{MS - MS} & 3.1 \times 10^6 /(3.1 \times 10^6) & 12.0/(12.0)   \\
            \mbox{MS - He}  & 2.3 \times 10^5 /(6.6 \times 10^3 ) & 19.0 /(20.4)\\
            \mbox{MS - G}  & 1.3 \times 10^4 /(1.2 \times 10^4) & 11.2 /(11.2)\\
            \mbox{He - He} & 1.2 \times 10^5 /(3.2 \times 10^4) & 10.4/(8.7)  \\
            \mbox{CO - MS}  & 1.3 \times 10^5 /(1.2 \times 10^5) & 14.2/(14.4) \\ 
            \mbox{CO - G} & 2.4 \times 10^3 /(2.4 \times 10^3) & 5.0 /(4.9)  \\ 
            \mbox{CO - He} & 8.0 \times 10^5/(2.1 \times 10^5) & 17.0 /(5.9)   \\
            \hline 
            \hline
            \multicolumn{3}{c}{\mbox{Disrupted systems}} \\
            \hline
            \hline
            \mbox{BH - BH} & 1.0 \times 10^6 /(1.0 \times 10^6) & 10.1/(10.1) \\
            \mbox{BH - NS} & 2.9 \times 10^6 /(2.8 \times 10^6) & 9.8/(9.7) \\ 
            \mbox{BH - WD} & 1.3 \times 10^5 /(1.3 \times 10^5)& 8.6/(8.6)  \\
            \mbox{BH - MS} & 2.1 \times 10^4 /(2.1 \times 10^4)& 8.4/(8.4)  \\
            \hline
            \hline
            \mbox{Total}:& 1.7 \times 10^7 /(1.6 \times 10^7) & 11.7/(11.3)   \\
            \noalign{\smallskip}
            \hline
            \hline
         \end{array}
     $$ 
   {Values for both models A and B given in order: A/(B). Note that in section with compact mergers in row BH-BH the number refers to number of single black holes that where created. In section with disrupted systems the number in row BH-BH informs about total number of single black holes after disruption (not the number of systems).}
   \end{table}
   
Average black hole mass in binary system is $\sim$ 17 M$_\odot$. The majority of binary systems with BH in bulge are BH-BH systems (80 \% of binary BHs). 
BHs in binary systems with non compact companion is a small fraction, about 3 \% of all bulge binary BHs.

   \begin{table}
      \caption[]{Binary systems with black holes in bulge\label{tab:binary}}
     $$ 
         \begin{array}{lrrr}
            \hline
            \hline
            \noalign{\smallskip}
            \mbox{Object type} & \mbox{Number} & \bar{\mbox{M}}_{{BH}} [\mbox{M}_\odot] & \bar{\mbox{M}}_{{Comp}} [\mbox{M}_\odot]\\
            \noalign{\smallskip}
            \hline
            \hline
            \noalign{\smallskip}
            \multicolumn{4}{c}{\mbox{Double compact object system}}    \\
            \hline
            \hline
            \mbox{BH - BH} & 7.6 \times 10^5 /(7.0 \times 10^5) & 17.1/(17.4)  &  - \\
            \mbox{BH - NS} & 3.1 \times 10^4 /(2.5 \times 10^4) & 11.8/(12.2) &  1.3/(1.3) \\
            \mbox{BH - WD} & 2.3 \times 10^5 /(2.1 \times 10^5) & 11.7/(11.8) & 1.0/(1.0) \\
            \hline
            \hline
            \noalign{\smallskip}
            \multicolumn{4}{c}{\mbox{Other binary system}}    \\
            \hline
            \hline
            \mbox{BH - MS} & 3.2 \times 10^4/(3.2 \times 10^4)  &11.0/(11.0)  &  0.7/(0.7) \\
            \mbox{BH - G} & 1.5 \times 10^3/(1.5 \times 10^3)  &8.8/(8.8)  &  1.3/(1.3) \\
            \hline
            \hline
            \mbox{Total:} & 1.0 \times 10^6 /(9.6 \times 10^5) & 15.7/ (15.9) & 1.0/ (1.0)  \\
            \hline
            \hline
         \end{array}
     $$ 
    Values for both models A and B given in order: A/(B). Note that the number in row with BH-BH systems refers to total number of black holes (not systems).
   \end{table}

\subsection{Galactic disk}\label{sec:disk}

About 80\% of Galactic BHs are in disk as in our simulation it is the most massive component ( see Sec. \ref{sec:sfr}). In both disk components, thin and thick, there are in total $\sim 1.0 \times 10^8$  single BHs with average mass of $\sim$ 14 M$_\odot$ and about $\sim$ 8 $\times 10^6$ black holes in binary systems with average mass of BH $\sim $19 M$_\odot$. The most massive black holes in Galactic disk are $\sim$ 100 M$_\odot$ and they are formed in coalescence of BH and and massive star (MS or He).

There are three main formation channels of single BHs in Galactic disk. Around 45 \% of single BHs in thin disk and in thick disk are final remnants of massive single star evolution. Second important channel is merger of binary systems which leads to formation of 25-30 \% of disk single BHs. BHs from disrupted binaries are about 20 \% of all disk single BHs with average masses of $\sim$ 12 M$_\odot$.

   \begin{table}
       \caption[]{Single black holes in thin disk \label{tab:singthin disk}}
     $$ 
         \begin{array}{lrr}
            \hline
            \hline
            \noalign{\smallskip}
             \mbox{Object type} & \mbox{Number} & \bar{\mbox{M}}_{BH} [\mbox{M}_\odot]\\
            \noalign{\smallskip}
            \hline
            \hline
            \noalign{\smallskip}
            \multicolumn{3}{c}{\mbox{Single stars}}    \\
            \hline
            \hline
            \mbox{Single star} & 4.2 \times 10^7 /(4.2 \times 10^7) & 14.3 /(14.3)\\
            \hline
            \hline
            \multicolumn{3}{c}{\mbox{Mergers}}    \\
            \hline
            \hline
            \mbox{BH - BH} & 1.5 \times 10^6 /(1.2 \times 10^5) & 29.8/(23.3) \\
            \mbox{BH - NS} & 1.1 \times 10^5 /(2.4 \times 10^4) & 10.8/(11.9) \\
            \mbox{NS - NS} & 3.3 \times 10^4 /(2.6 \times 10^3) & 2.8 /(2.6) \\
            \mbox{NS - WD} & 1.5 \times 10^4 /(3.5 \times 10^3) & 2.5 /(2.9) \\
            \mbox{MS - MS} &  1.6 \times 10^7 /(1.6 \times 10^7) & 15.3/(15.3) \\
            \mbox{MS - He}  & 1.7 \times 10^6 /(5.2 \times 10^4) & 20.2/(20.6) \\
            \mbox{MS - G}  & 4.4 \times 10^4 /(4.9 \times 10^4) & 17.7/(16.0) \\
            \mbox{He - He}  & 6.7 \times 10^5 /(2.3 \times 10^5) & 10.5/(8.8)
            \\
            \mbox{CO - MS}  & 8.7 \times 10^5 /(8.1 \times 10^5) & 16.1/(16.5)
            \\
            \mbox{CO - G}  & 1.4 \times 10^4 /(1.2 \times 10^4) & 5.2/(5.8) \\
            \mbox{CO - He} & 5.5 \times 10^6 /(1.6 \times 10^6) & 17.7/(5.7) \\
            \hline
            \hline
            \multicolumn{3}{c}{\mbox{Disrupted systems}}    \\
            \hline
            \hline
            \mbox{BH - BH} & 1.1 \times 10^7 /(1.1 \times 10^7) & 11.5(11.5)\\
            \mbox{BH - NS} & 1.4 \times 10^7 /(1.4 \times 10^7) & 11.6(11.6)
            \\
            \mbox{BH - WD} & 9.8 \times 10^5 /(9.8 \times 10^5)& 9.1(9.1)
            \\
            \mbox{BH - MS} & 1.5 \times 10^5 /(1.5 \times 10^5)& 9.0(9.0)
            \\
            \hline
            \hline
            \mbox{Total:} & 9.5 \times 10^7 /(8.8 \times 10^7) & 14.0 (13.5) \\
            \hline
            \hline
         \end{array}
     $$ 
    {Values for both models A and B given in order: A/(B). Note that in section with compact mergers in row BH-BH the number refers to number of single black holes that where created. In section with disrupted systems the number in row BH-BH informs about total number of single black holes after disruption (not the number of systems).}
   \end{table}
   \begin{table}
       \caption[]{Single black holes in thick disk \label{tab:singthick disk}}
     $$ 
         \begin{array}{lrr}
            \hline
            \hline
            \noalign{\smallskip}
             \mbox{Object type} & \mbox{Number} & \bar{\mbox{M}}_{BH} [\mbox{M}_\odot]\\
            \noalign{\smallskip}
            \hline
            \hline
            \noalign{\smallskip}
            \multicolumn{3}{c}{\mbox{Single stars}}    \\
            \hline
            \hline
            \mbox{Single star} & 4.3 \times 10^6 /(4.3 \times 10^6) & 17.0 /(17.0)\\
            \hline 
            \hline
            \multicolumn{3}{c}{\mbox{Mergers}}    \\
            \hline
            \hline
            \mbox{BH - BH} & 2.5 \times 10^5 /(2.9 \times 10^4)  & 30.2/(17.8) \\
            \mbox{BH - NS} & 1.7 \times 10^4 /(3.9 \times 10^3)  & 11.8/(12.3) \\
            \mbox{NS - NS} & 8.3 \times 10^2  /(3.8 \times 10^2) & 2.7 /(2.6) \\
            \mbox{MS - MS} & 1.5 \times 10^6 /(1.5 \times 10^6) & 18.9/(18.9) \\
            \mbox{MS - He}  & 2.3 \times 10^5 /(3.3 \times 10^3) & 19.9/() \\
            \mbox{MS - G}  &  4.4 \times 10^3/(3.9 \times 10^3) & 17.7/(18.2) \\
            \mbox{He - He}  & 1.3 \times 10^5 /(4.5 \times 10^4) & 10.0/(8.5) \\
            \mbox{CO - MS}  & 1.0 \times 10^5 /(1.0 \times 10^5) & 17.6/(19.7) \\
            \mbox{CO - G} & 2.0 \times 10^3 /(2.2 \times 10^3) & 4.7/(5.4) \\
            \mbox{CO - He} & 7.4 \times 10^5 /(2.3 \times 10^5) & 19.1/(6.1) \\
            \hline
            \hline
            \multicolumn{3}{c}{\mbox{Disrupted systems}}    \\
            \hline
            \hline
            \mbox{BH - BH} & 8.3 \times 10^5 /(8.3 \times 10^5) & 12.8/(12.8)\\
            \mbox{BH - NS} & 1.3 \times 10^6 /(1.3 \times 10^6)& 13.8/(13.8)
            \\
            \mbox{BH - WD} & 8.8 \times 10^4 /(8.8 \times 10^4)& 10.3/(10.3)
            \\
            \mbox{BH - MS} & 9.0 \times 10^3 /(9.0 \times 10^4)& 10.3/(10.3)
            \\
            \hline
            \hline
            \mbox{Total:} & 9.5 \times 10^6 /(8.5 \times 10^7) & 17.0 /(16.3)\\
            \hline
            \hline
         \end{array}
     $$ 
    {Values for both models A and B given in order: A/(B). Note that in section with compact mergers in row BH-BH the number refers to number of single black holes that where created. In section with disrupted systems the number in row BH-BH informs about total number of single black holes after disruption (not the number of systems).}
   \end{table}
   
Majority of binary BHs in disk ($\sim$ 80 \% of BH in binaries) are in BH-BH configuration. The average mass of BH in binary system is about $\sim$ 19 M$_\odot$. The fraction of BHs with unenvolved companion is small, less than $2 \%$ of all disk binary BHs. 

   \begin{table}
      \caption[]{Binary systems with black holes in thin disk\label{tab:binaryThinDisk}}
     $$ 
         \begin{array}{lrrr}
            \hline
            \hline
            \noalign{\smallskip}
            \mbox{Object type} & \mbox{Number} & \bar{\mbox{M}}_{BH} [\mbox{M}_\odot] & \bar{\mbox{M}}_{Comp} [\mbox{M}_\odot]\\
            \noalign{\smallskip}
            \hline
            \hline
            \noalign{\smallskip}
            \multicolumn{4}{c}{\mbox{Double compact object system}}    \\
            \hline
            \hline
            \mbox{BH - BH} &  5.5 \times 10^6/ (4.6 \times 10^6) & 19.6/(20.9)  &  - \\
            \mbox{BH - NS} & 2.1 \times 10^5 /(1.2 \times 10^5) & 12.8/(15.0) &  1.3/(1.3) \\
            \mbox{BH - WD} & 1.2 \times 10^6 /(1.0 \times 10^6) & 14.3/(14.8) & 1.0/(0.9) \\
            \hline
            \hline
            \noalign{\smallskip}
            \multicolumn{4}{c}{\mbox{Other binary system}}    \\
            \hline
            \hline
            \mbox{BH - MS} & 1.2 \times 10^5 /(1.2 \times 10^5) & 16.4/(16.4) & 0.6/(0.6) \\
            \mbox{BH - G} & 6.6 \times 10^3 /(6.6 \times 10^3) & 14.2/(14.2) & 0.9/(0.9) \\
            \mbox{BH - HG} & 1.6 \times 10^3 /(1.6 \times 10^3) & 15.4/(15.4) & 1.3/(1.3) \\
            \hline
            \hline
            \mbox{Total:} & 7.0 \times 10^6 /(5.9 \times 10^6) & 18.5/(19.6) & 1.0/(0.9)  \\
            \hline
            \hline
         \end{array}
     $$ 
    Values for both models A and B given in order: A/(B). Note that the number in row with BH-BH systems refers to total number of black holes (not systems).
   \end{table}

   \begin{table}
      \caption[]{Binary systems with black holes in thick disk\label{tab:binaryThickDisk}}
     $$ 
         \begin{array}{lrrr}
            \hline
            \hline
            \noalign{\smallskip}
            \mbox{Object type} & \mbox{Number} & \bar{\mbox{M}}_{BH} [\mbox{M}_\odot] & \bar{\mbox{M}}_{Comp} [\mbox{M}_\odot]\\
            \noalign{\smallskip}
            \hline
            \hline
            \noalign{\smallskip}
            \multicolumn{4}{c}{\mbox{Double compact object system}}    \\
            \hline
            \hline
            \mbox{BH - BH} &  6.5 \times 10^5 /(5.3 \times 10^5)& 21.7/(23.1) & - \\
            \mbox{BH - NS} & 2.0 \times 10^4 /(1.4 \times 10^4) &15.2/(15.9)& 1.3/(1.3) \\
            \mbox{BH - WD} & 1.6 \times 10^5 /(1.4\times 10^5) & 14.7/(15.2) & 0.9/(0.9) \\
            \hline
            \hline
            \multicolumn{4}{c}{\mbox{Other binary systems}} \\
            \hline
            \hline
            \mbox{BH - MS} &  1.2 \times 10^4 /(1.2 \times 10^4)& 18.1/(18.1) & 0.5/(0.5) \\
            \mbox{BH - G} &  2.9 \times 10^2 /(2.9 \times 10^2)& 17.9/(17.9) & 0.6/(0.6) \\
            \hline
            \hline
            \noalign{\smallskip}
            \mbox{Total:} & 8.3 \times 10^5 /(6.9 \times 10^5) & 20.2 /(21.2) & 0.9 /(0.9) \\
            \hline
            \hline
         \end{array}
     $$ 
    Values for both models A and B given in order: A/(B). Note that the number in row with BH-BH systems refers to total number of black holes (not systems).
   \end{table}

\subsection{Galactic halo} \label{sec:halo}
Galactic halo is the least massive component in our simulation as we considered only its stellar mass (Sec. \ref{sec:sfr}). Majority of halo mass are not stars, so we do not consider this mass in stellar origin BH formation. The total number of black holes in halo is only 4 \% of all Galactic BH population, it contains $\sim$ 4 $\times 10^6$ single black holes with average mass $\sim$ 20 M$_\odot$ and about $\sim$ 5 $\times 10^5$ black holes in binary systems with average BH mass $\sim$ 24 M$_\odot$. \\  Due to the low metallicity of stellar populations, the most massive black hole in simulations, with mass of 113 M$_\odot$, is formed in halo. It originates from a merger of BH and He star. 
Formation channels of single black holes in halo are the same as in disk and bulge. Over 50\% of single BHs are remnants of single massive star evolution while 20-25\% formed in binary system mergers (mainly MS+MS, MS+He and WD+He). BHs from disrupted binary systems are $\sim$ 15\% of halo single BHs with their average mass of about 17 M$_{\odot}$.

\begin{table}
      \caption[]{Single black holes in halo\label{tab:singleHalo}}
     $$ 
            \begin{array}{lrr}
            \hline
            \hline
            \noalign{\smallskip}
             \mbox{Object type} & \mbox{Number} & \bar{M}_{BH} [M_\odot]\\
            \noalign{\smallskip}
            \hline
            \hline
            \noalign{\smallskip}
            \multicolumn{3}{c}{\mbox{Single stars}}    \\
            \hline
            \hline
            \mbox{Single star} & 2.1 \times 10^6 /(2.1 \times 10^6)  & 19.0/(19.0)    \\
            \hline
            \hline
            \noalign{\smallskip}
            \multicolumn{3}{c}{\mbox{Compact mergers}}    \\
            \hline
            \hline
            \mbox{BH - BH} & 6.2 \times 10^4 /(3.9 \times 10^4) & 41.0/(42.5)  \\
            \mbox{BH - NS} & 7.4 \times 10^2 /(7.4 \times 10^2) & 11.7/(11.7) \\
            \mbox{NS - NS} & 2.5 \times 10^2 /(4.8 \times 10^1) & 2.5/(2.5) \\
            \hline
            \hline
            \noalign{\smallskip}
            \multicolumn{3}{c}{\mbox{Mergers}}    \\
            \hline
            \hline
            \mbox{MS - MS} & 4.7 \times 10^5 /(4.7 \times 10^5) & 22.9/(22.9)\\
            \mbox{MS - He}  & 1.3 \times 10^5 /(3.7 \times 10^4) & 26.6/(30.3)\\
            \mbox{MS - G}  & 1.2 \times 10^3 /(1.3 \times 10^3) & 23.2/(23.6)\\
            \mbox{He - He}  & 2.3 \times 10^4 /(2.0 \times 10^4) & 13.5/(11.7) \\
            \mbox{CO - MS}  & 6.9 \times 10^4 /(6.9 \times 10^4) & 14.6/(14.6) \\
            \mbox{CO - G} & 2.2 \times 10^2 /(2.0 \times 10^2) & 11.3/(13.0) \\
            \mbox{CO - He} & 4.2 \times 10^5 /(2.0 \times 10^5) & 24.8/(11.0) \\
            \noalign{\smallskip}
            \hline
            \hline
            \multicolumn{3}{c}{\mbox{Disrupted systems}}    \\
            \hline
            \hline
            \mbox{BH - BH} & 2.9 \times 10^5 /(1.2 \times 10^5) & 16.8/(15.0) \\
            \mbox{BH - NS} & 5.2 \times 10^5 /(2.5 \times 10^5) & 17.7/(16.0) \\
            \mbox{BH - WD} & 3.8 \times 10^4 /(1.8 \times 10^4) & 13.1/(12.7) \\
            \mbox{BH - MS} & 4.7 \times 10^3 /(4.7 \times 10^3) & 13.5/(13.5) \\
            \hline 
            \hline
            \mbox{Total:} & 3.7 \times 10^6 /(3.2 \times 10^6) & 21.0 /(19.9) \\
            \hline
            \hline
         \end{array}
     $$ 
     {Values for both models A and B given in order: A(B). Note that in section with compact mergers in row BH-BH the number refers to number of single black holes that where created. In section with disrupted systems the number in row BH-BH informs about total number of single black holes after disruption (not the number of systems).}
   \end{table}

Binary black holes in halo are also mainly in BH-BH binaries (in $\sim $80\% of binary BHs). The fraction of BHs with non compact companion in halo is $\sim$ 1.5 \%.

   \begin{table}
      \caption[]{Binary systems with black holes in halo\label{tab:binaryHalo}}
     $$ 
         \begin{array}{lrrr}
            \hline
            \hline
            \noalign{\smallskip}
            \mbox{Object type} & \mbox{Number} & \bar{\mbox{M}}_{BH}
            [\mbox{M}_\odot] & \bar{\mbox{M}}_{Comp} [\mbox{M}_\odot]\\
            \noalign{\smallskip}
            \hline
            \hline
            \noalign{\smallskip}
            \multicolumn{4}{c}{\mbox{Double compact object system}}    \\
            \hline
            \hline
            \noalign{\smallskip}
            \mbox{BH - BH} & 3.7 \times 10^5 /(3.7 \times 10^5) & 25.0/(25.0) & -     \\
            \mbox{BH - NS} & 1.1 \times 10^4 /(1.1 \times 10^4) & 21.4/(21.4) &  1.3/(1.3)  \\
            \mbox{BH - WD} & 7.0 \times 10^4 /(7.0 \times 10^4) & 20.5/(20.5) & 1.0/(1.0) \\
            \hline
            \hline
            \multicolumn{4}{c}{\mbox{Other binary systems}} \\
            \hline
            \hline
            \mbox{BH - MS} & 7.7 \times 10^3 /(7.7 \times 10^3) & 21.6/(21.6) & 1.0/(1.0) \\
            \mbox{BH - G} & 1.8 \times 10^2 /(1.8 \times 10^2) & 22.0/(22.0) & 0.6/(0.6) \\
            \hline
            \hline
            \mbox{Total:} &  4.6 \times 10^5 /(4.6 \times 10^5) & 24.2/(24.2) & 1.0/(1.0) \\
            \noalign{\smallskip}
            \hline
            \hline
         \end{array}
     $$ 
     {Values for both models A and B given in order: A(B). Note that the number in row with BH-BH systems refers to total number of black holes (not systems).}
   \end{table}

 \subsection{Dark matter in stellar origin BHs} \label{darmatter}
 
There have been several microlensing surveys towards halo in order to search for and constrain the amount of massive compact halo objects (MACHOs) which may constitute part of dark matter \citep{2001ApJS..136..439A}. The presence of MACHOs with mass over 20 M$_\odot$ can not be excluded by observations \citep{2007A&A...469..387T,2011MNRAS.416.2949W} as current microlensing surveys do not cover long enough timescale to detect such massive objects. On the other hand, the existence of wide binary systems in halo indicates the absence of MACHOs with masses larger than $\sim$ 100 M$_\odot$. Observed binaries would likely get disrupted in interaction with objects with such a large mass \citep{2004ApJ...601..311Y,2014ApJ...790..159M}. 
Those two restrictions give us the recent upper and lower limit on MACHOs mass range which existence cannot be excluded by current observational surveys \citep{2016PhRvL.116t1301B}. 

We calculated fraction of mass in Galactic halo hidden in the form of stellar origin black holes in the mass range of 20-100 M$_\odot$. The total mass of such BHs is $\sim$ 6.2 $\times 10^7$ M$_\odot$, which could constitute only $\sim$ 0.006 \% of total Galactic dark matter mass which is of the order of $\sim 10^{12}$M$_\odot$ \citep{2015MNRAS.453..377W,2018A&A...616L...9M,2019MNRAS.487L..72G}.

\subsection{Mass distribution} \label{sec:mass_distribution}
   
   In Figures \ref{mass_distributionb} and \ref{mass_distributionad} we present distribution of single and binary BHs masses for two considered evolutionary models A and B  and three Galactic components: bulge, disk and halo. Mass distribution for single and binary BHs is very different. Average mass of a BH in a binary system $\sim 19$ M$_\odot$ is larger than for single BHs with average mass of $\sim 14$ M$_\odot$. The difference in average mass of single and binary BHs is mainly due to our adopted natal kick distribution, in the less massive remnant (BH/NS) the higher velocity it gets. This results in disruption of many low mass binaries during BH/NS formation. Note that in all Galactic components mean BH mass in binary BH-BH system is larger than a mass of BH from distrupted BH-BH systems (see Tab: \ref{tab:singlebulge}, \ref{tab:binary}, \ref{tab:singthin disk}, \ref{tab:singthick disk}, \ref{tab:binaryThinDisk},  \ref{tab:binaryThickDisk}, \ref{tab:singleHalo}, \ref{tab:binaryHalo}). Also because of assumed natal kick distribution, BHs in binary system are mainly in BH-BH systems (80 \% of Galactic binary BHs).
   
    The distributions of single BH masses in all Galactic components (disk, bulge and halo) have a peak near 10-15 M$_{\odot}$ and above that mass the number of BHs systematically decreases and reaches zero in different mass limits depending on Galactic component (note log scale). In halo, the decrease of the number of BHs along with mass is less steep than in bulge or disk. The average single BH mass in halo ($\sim$ 17 M$_\odot$) is larger than in other components (bulge and disk $\sim$ 12-13 M$_\odot$). The occurrence of more massive BHs is due to the lower stellar metallicity \citep{2010ApJ...714.1217B} which is associated with less weight loss from massive stars due to stellar winds. 
   The range of possible single BH masses ($\sim$ 2.5-113 M$_\odot$) is wide as significant number of single BHs originate from binary system mergers ($\sim$ 20-30 \%), which could be both low mass (the final BH mass close to max. NS limit) and high mass. Average mass of BHs of all merger types is similar to other formation channels. However, note that BHs from mergers widen the range of possible BH masses above PPSN limit and fill the first and second mass gap (merger of massive black hole with its binary companion). The first mass gap \citep{2010Natur.467.1081D,2013Sci...340..448A,Swihart_2017} between 3-5 M$_\odot$.is filled by black hole masses formed mainly in coalescence of WD and He/MS stars. We assume that merger of WD and a massive star leads to the collapse to a BH/NS.   
   As we mentioned before, single BHs are hard to detect so the presence of black holes in the mass gap is not in tension with observations \citep{2019arXiv190407789W}. The second mass gap is filled mainly by mergers of BH+He, BH+MS stars and BH-BH. In both bulge and disk the largest achieved BH mass is $\sim$ 100 M$_\odot$ while in halo it is over 110 M$_\odot$.  
   
   To check the impact of the adopted criterion for estimating the masses of objects formed in mergers (on for example formation of objects in the first and second mass gaps), we tested two alternative simplified methods. In the first method we set factors which define fraction of accreted mass of the merging companion (see $f_{MS}, f_{G}, f_{He}$ Sec. \ref{sec:mergers}) to zero so we do not allow for accretion from the second object. In the second method, the factors were set to 0.5 so we assumed accretion of a half the mass of the companion. Note that for double compact object mergers we made an exception and took a sum of merging objects without any mass loss. The distribution of BH masses for the non-accretion and half-accretion cases, for two CE models A and B are shown in the Figures \ref{mass_distributionb00}, \ref{mass_distributionad00}, \ref{mass_distributionb05}, \ref{mass_distributionad05}.
   
   In general the use of different methods did not have large impact on the average single black hole masses in Galactic components, especially considering evolutionary model B. However, one may notice a significant reduction in the number of low mass black holes below 5 M$_\odot$(first mass gap) in the results for non-accretion model (f=0.0, Fig. \ref{mass_distributionb00} and \ref{mass_distributionad00}). In three tested models (standard, f=0.0 and f=0.5) both mass gaps are filled by the products of mergers. However the number of BHs in first and the second mass gap strongly depends on the assumed accretion factor f. Note that in the non-accretion model objects in the second mass gap are only products of double compact object mergers while in the other two models, BHs with masses over PPSN limit could have formed for example in BH and MS or BH and HE star coalescence. The average single BH masses for the three tested mass estimation models are given in the Table \ref{tab:merger_masses}.  
   Due to the reduction of accretion of matter during merger also the total number of single BHs is the lower the smaller accretion factor f is assumed. The total number of single BHs was reduced by about 20 \% for non-accretion model and about 1 \% for the model with half-accretion in comparison to our standard model described in Section \ref{sec:mergers}.
   
      \begin{table} 
      \caption[]{Average single BH masses in Galactic components: bulge, disk and halo. Results for two models A and B in the order A(B), see Sec. \ref{sec:models}. The results for three methods of mass estimation of objects formed in mergers. The $\overline{M}$ is the standard method described in Sec. \ref{sec:mergers}. $ \overline{M}_{f = 0}$ and  $\overline{M}_{f = 0.5}$ are models with zero and half accretion from the merging companion (factors $f_{MS}, f_{G}, f_{He}$ are 0 or 0.5) }\label{tab:merger_masses}
     $$ 
         \begin{array}{lccc}
            \hline
            \hline
            \noalign{\smallskip}
           \mbox{} & \overline{M} [M_{\odot}] & \overline{M}_{f= 0} [M_{\odot}] & \overline{M}_{f = 0.5} [M_{\odot}] \\
            \noalign{\smallskip}
            \hline
            \hline
            \mbox{Bulge} & 11.7(11.3)  & 11.4(11.3) & 11.5(11.3)\\
            \mbox{Disk} & 14.2(13.7) & 13.7(13.7) & 14.0(13.6) \\
            \mbox{Halo} & 21.0(19.9) & 20.1(20.1) & 20.3(19.6) \\
            \hline
            \hline
         \end{array}
     $$ 
   \end{table}

   The range of possible binary BH masses is narrower than for single black holes. The upper limit on binary BHs mass ($\sim$ 50-60 M $_\odot$) is similar in all Galactic components: disk, bulge and halo and it is a result of two physical processes: stellar winds \citep{2010ApJ...714.1217B}) and pair-instability limit \citep{2019arXiv190111136L,2017ApJ...836..244W}. The upper limit is consistent with the second observational mass gap. Also the first mass gap is reconstructed for binary BHs due to the adopted rapid SN model \citep{Fryer1204}. However, there is a narrow, isolated peak near a range of 2.5-3 M$_\odot$ (close to max. mass of NS). The peak is made of black holes that formed in accretion of matter on the massive neutron star from its binary companion. The fraction of BHs with unevolved companion (e.g. main sequence or giant star) is small, a few percent of all binary BHs. The companion is usually a low massive star M<1 M$_\odot$. More massive stars are less frequent, especially in older stellar populations as their evolution time is shorter than most population ages. Even if the NS interacts with the more massive companion, the mass transfer on the NS/BH during short-lived CE event is related to the Bondi-Hoyle accretion which in our model is rather inefficient (see Sec. \ref{sec:method}).

\subsection{Separation and eccentricity of BH binary systems} 
In our database we include orbital parameters (separation and eccentricity) of all binary systems which contain a black hole. Distribution of systems separation in Galactic components (bulge, disk, halo), for CE model A (top) and B (bottom) is plotted in Figure \ref{separation_distribution} (log-log scale) and covers wide range from a few R$_\odot$ to $10^8$ R$_\odot$. Average separation is different in Galactic components, in bulge it is $\sim 8.8 \times 10^4 (9.4 \times 10^4)$ R$_\odot$, in disk $\sim 6.4 \times 10^4 (7.4 \times 10^4)$ R$_\odot$ while in halo $\sim 3.9 \times 10^4 (3.9 \times 10^4)$ R$_\odot$.

Distribution of BH systems eccentricities in Galactic components is shown in Figure \ref{eccentricity_distribution}, for model A (top) and model B (bottom). It is dominated by low eccentricities, with high peak near 0-0.1. The average eccentricity of BH binary systems is similar in all Galactic components and is in the range of [$0.15-0.18$]. 

Note that components in wide binary systems did not exchange mass during the evolution and the orbit was only affected by wind mass loss, magnetic braking or gravitational waves emission. \color{black} 

\subsection{Velocity} \label{sec:velocity}

In Figures \ref{velocity_distributionb} and \ref{velocity_distributionad} we plotted a distribution of a single and a binary black hole velocities for the two considered models A and B (see Sec.\ref{sec:models}) and three Galactic components: bulge, disk and halo.
With the red, dashed line we marked the velocity equal 580 km/s which is an estimated value of the local Galactic escape speed at the Sun's position \citep{2018A&A...616L...9M}. Escape speed depends on the location in the Milky Way and can take values from a range of $\sim$ 550-650 kms$^{-1}$.

Average value for both single and binary BHs is similar in given Galactic component as BHs in general do not reach high speeds as a result of binary or single evolution. Majority of BH velocities are close to the values which we adopted from approximated form of rotation curve for a given component (\ref{sec:method}). Very high speeds of single BHs (over 1000km/s) are rare cases (note log scale).
 However, there is a difference in the range of possible speeds of single and binary BHs. Single BHs may achieve speeds from 0 to even $\sim$ 1700 km/s in model A and to 1100 kms$^{-1}$ in model B. The maximum speed of BH in binary system is $\sim$ 700 kms$^{-1}$. This is a result of assumed natal kick distribution (see Sec, \ref{sec:method}) . NS and low massive BHs gets higher natal kicks, inversely proportional to remnant mass. Those systems often get disrupted. BHs from disrupted binaries may achieve high velocities and are classified as single BHs. Black holes in binary systems are often more massive and get lower natal kicks (remain bound).
 Sporadic cases of very high speeds in model A are low massive BHs from close binaries which got disrupted after BH/NS formation which pass CE phase with HG donor. 

In the Table \ref{tab:velocities} we present average single and binary BH speeds for two models A and B and different Galactic bulge, disk and halo.

   \begin{table} 
      \caption[]{Average values of BH speeds in Galactic components: bulge, disk and halo. Results for two models A and B in the order A(B), see Sec. \ref{sec:models}} \label{tab:velocities}
     $$ 
         \begin{array}{lccc}
            \hline
            \hline
            \noalign{\smallskip}
           \mbox{} & \mbox{Bulge [km/s]} & \mbox{Disk [km/s]} & \mbox{Halo [km/s]} \\
            \noalign{\smallskip}
            \hline
            \hline
            \mbox{Single BH} & 140(140) & 238(239) & 242(246)\\
            \mbox{Binary BH} & 112(111) & 221(221) & 220(220) \\
            \hline
            \hline
         \end{array}
     $$ 
   \end{table}
 
We calculated that $\sim$ 5\% of single BHs ($\sim 6 \times 10^6$ ) and less than 0.0001\% of binary BHs ($\sim 10$) have velocities greater than 550 km/s, the lowest escape velocity from Milky Way \citep{2018A&A...616L...9M}. This gives us an upper limit on the fraction of BHs that could escape form the Galaxy in our physical model. Note that if we would adopt natal kick model with no fallback parameter the fraction of BHs with very high velocities would increase.\color{black}

\subsection{BH binaries with non compact companion}

We present parameters of our synthetic population of Galactic BHs with non evolved binary companion, which could be MS, G or He star. This fraction of BH systems might be especially interesting for BH hunters as they are more likely to be detected due the presence of visible companion. However they seem to consist only a small fraction (less than 1\% of whole Galactic BHs). \\
In the Fig. \ref{fig:BH_binaries_mass} and \ref{fig:BH_binaries_masB} we plot the mass distribution of BHs (top) and their companions (bottom) for two considered CE models A and B and three Galactic components (bulge, disk and halo). Mass distribution is similar for models A and B while both BHs and their companions mass distribution varies depending on given Galactic component. Mean masses of BHs in Galactic components are: $\sim 16$M$_\odot$ in disk, $\sim 11$M$_\odot$ in bulge, $\sim 21$M$_\odot$ in halo while mean masses of companion stars are $\sim 3.5$M$\odot$ in disk, $\sim 2.0 $M$\odot$ in bulge and $\sim 0.5$ M$\odot$ in halo. The differences in BH masses are mainly the effect of the metallicity, which is very low in halo comparing with disk and bulge. The difference in the companion masses distribution is the result of age and star formation history of given component. We assumed that stellar populations in halo formed 10-12 Gyr ago so only low mass stars haven't finished their evolution yet. On the other hand we assumed presence of many young stars in disk and also some young populations in bulge so binaries with more massive stars are still possible to be found there.  
We plot orbital parameters (separation and eccentricity) of the systems. Distribution of separations is shown in Fig. \ref{fig:BH_binaries_separation} while eccentricities in Fig. \ref{fig:BH_binaries_eccentricity}. Distribution of separations covers wide range from a few R$_\odot$ to $10^7$ R$_\odot$ and seems to be approximately uniform on log-log scale.
The eccentricities of the systems are rather low, there is a peak in the number of systems with eccentricities 0-0.1 while the average of all Galactic components is about 0.25. In Figure \ref{fig:BH_binaries_e_a} we plotted diagram which illustrates the relationship between systems separations and eccentricities.

\subsection{Massive BHs from MS+He mergers}

Among our results we found some cases of MS+He star mergers which can produce a  star with a helium core mass below pair-instability limit with its total mass (core + envelope) as high as 60-90 M$_\odot$. We assumed the naked helium star is more compact than main sequence star and therefore the mass of the core after merger would not increase so total mass accreted from the MS companion will create the envelope.
It has been showed by \cite{2017ApJ...836..244W} that for an ultra-low metallicity and Population III stars the  PPSN/PSN  in-stability can be shifted to max. $\sim 70$ M$_\odot$ as such stars can keep massive H-rich envelopes.
Considering this limit we have removed from our database BHs formed in MS+He merger which are more massive than 70 M$_{\odot}$. 
However, single BHs with masses above that limit can be still created in black hole mergers e.g. BH+MS or BH+BH mergers. Similar scenario was considered recently by \cite{2019arXiv191204509T} in the context of LB-1 formation.

\subsection{Galactic merger rates} \label{sec:merger_rates}

\begin{figure}  
    \includegraphics[width=99 mm]{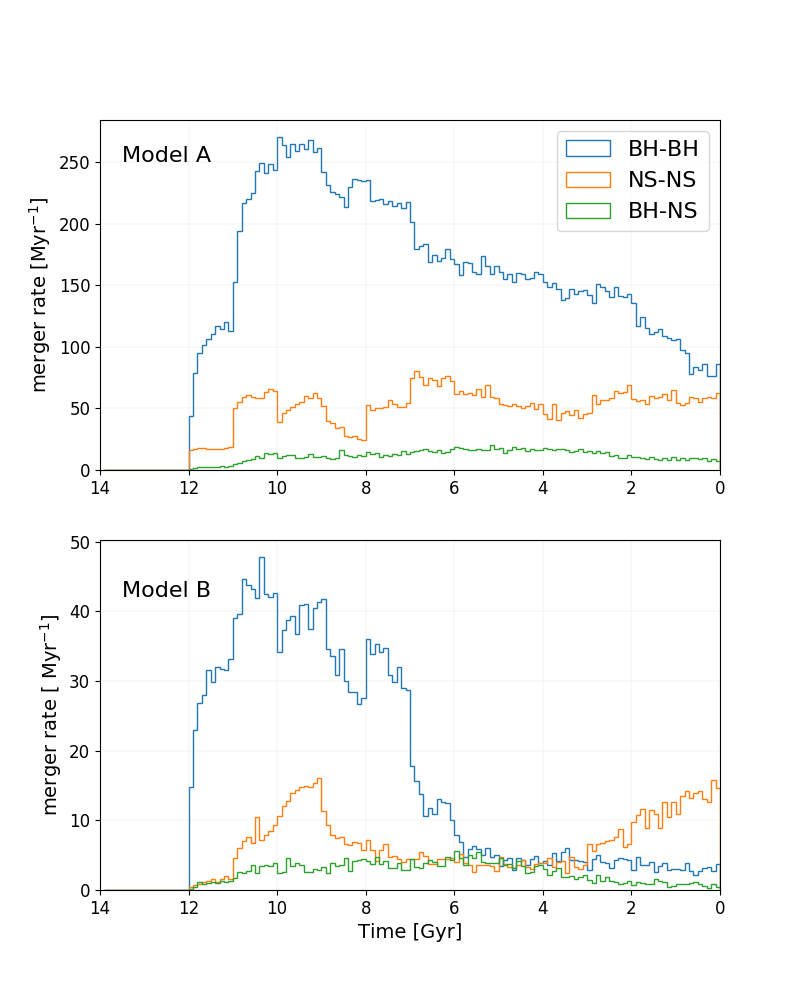} 
    \caption{Merger rates of double compact object binaries (BH-BH, BH-NS, NS-NS) as a function of time since Big Bang. Results for new Galactic SFR and chemical evolution model (Sec. \ref{sec:sfr}) and two evolutionary models A and B, see section \ref{sec:models}. Current time is 0.} 
    \label{fig:merger_rates}  
\end{figure}

\begin{figure}  
    \includegraphics[width=99 mm]{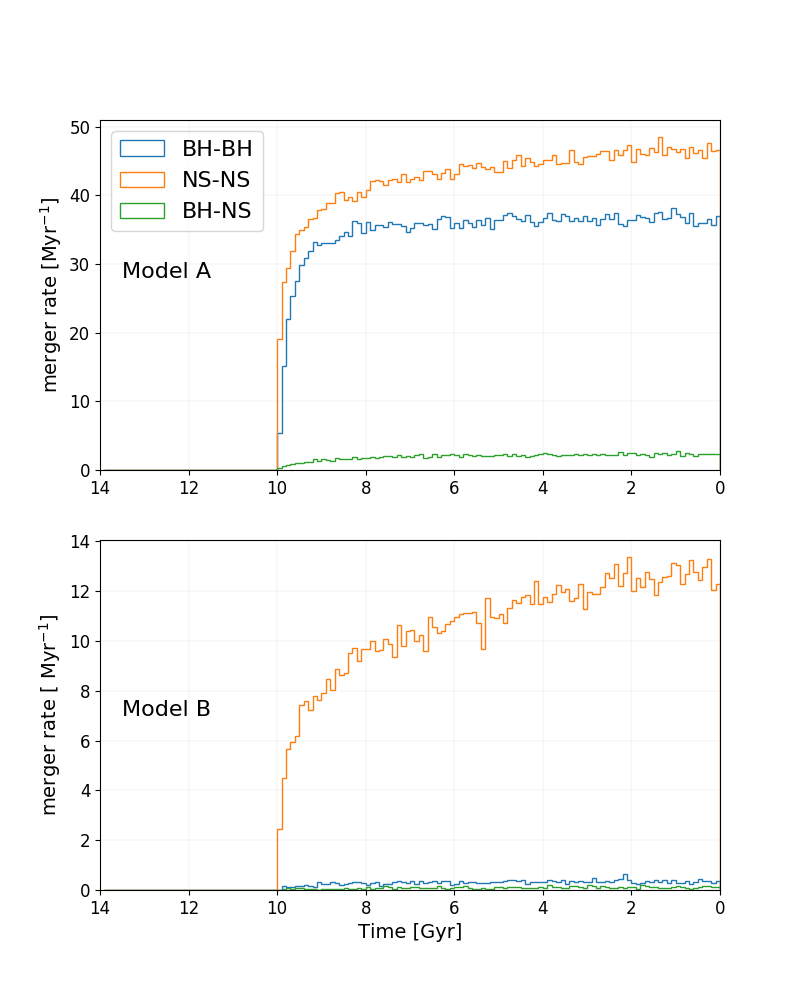} 
    \caption{ Merger rates of double compact object binaries (BH-BH, BH-NS, NS-NS) as a function of time since Big Bang. Results for old Galactic SFR and chemical evolution model and two evolutionary models A and B, see section \ref{sec:models}. Current time is 0.} 
    \label{fig:merger_rates2}  
\end{figure}

\begin{figure} 

    \includegraphics[width=95mm]{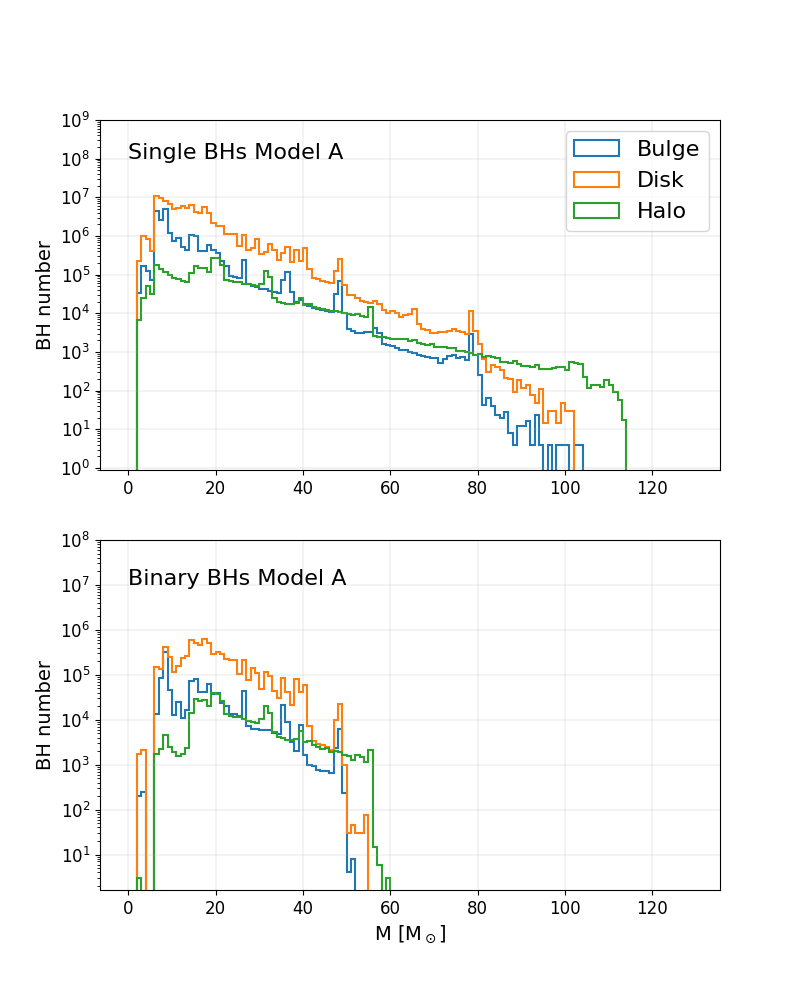} 
    \caption{Single (top) and binary (bottom) BHs mass distribution in three Galactic components for model A. In model A we allow binary system to survive CE phase with HG donor star. Mass of the objects formed in merger estimated as explained in \ref{sec:mergers}.} 
    \label{mass_distributionb}
\end{figure}

\begin{figure}    
    \includegraphics[width=95 mm]{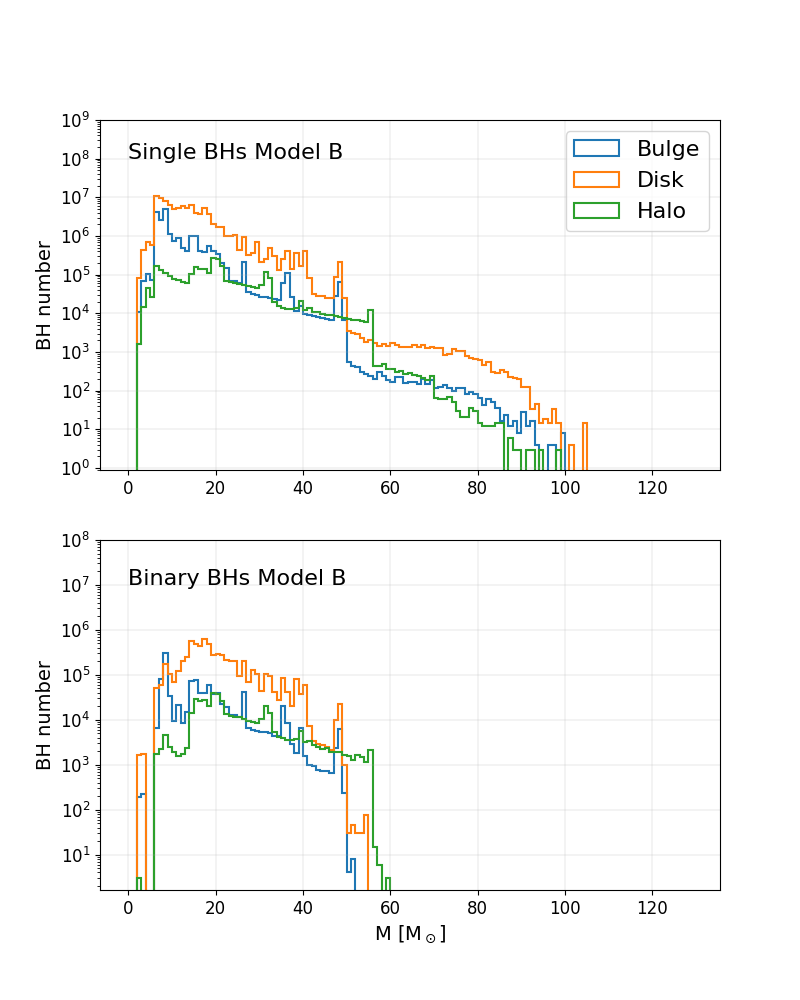} 
    \caption{Single (top) and binary (bottom) BHs mass distribution in three Galactic components for model B. In model B we assume that CE phase with HG star donor always leads to binary system merger. Mass of the objects formed in merger estimated as explained in \ref{sec:mergers}.} 
    \label{mass_distributionad}
\end{figure}

\begin{figure} 

    \includegraphics[width=95mm]{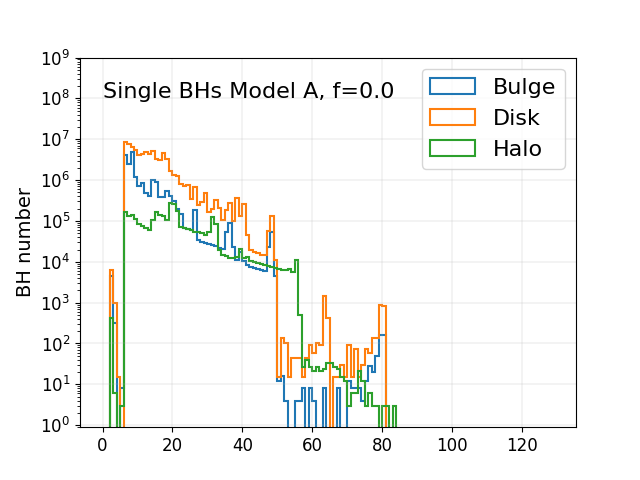} 
    \caption{Single BH mass distribution in three Galactic components for model A. In model A we allow binary system to survive CE phase with HG donor star. Mass of the objects formed in merger estimated as a lower limit assuming total mass loss from the second merging object (except for two compact object mergers). } 
    \label{mass_distributionb00}
\end{figure}

\begin{figure}    
    \includegraphics[width=95 mm]{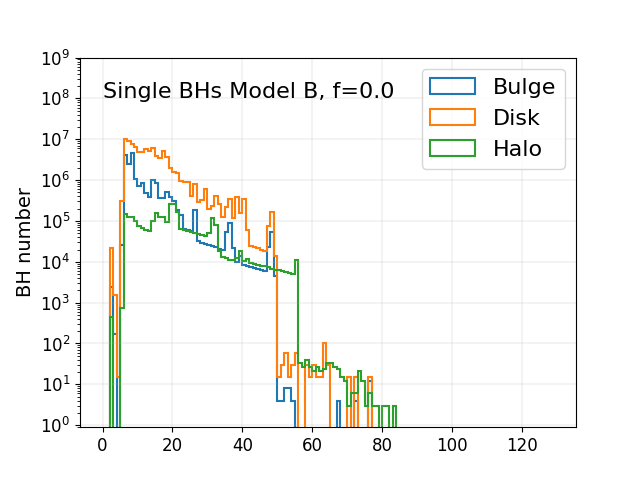} 
    \caption{Single BH mass distribution in three Galactic components for model B. In model B we assume that CE phase with HG star donor always leads to binary system merger. Mass of the objects formed in merger estimated as a lower limit assuming total mass loss from the second merging object (except two compact object mergers). } 
    \label{mass_distributionad00}
\end{figure}

\begin{figure} 

    \includegraphics[width=95mm]{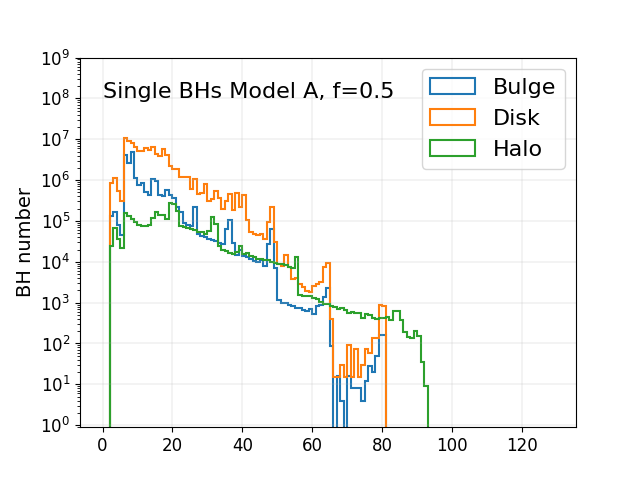} 
    \caption{Single BH mass distribution in three Galactic components for model A. In model A we allow binary system to survive CE phase with HG donor star. Mass of the objects formed in merger assuming 50 \% mass loss from the second merging object (except two compact object mergers).} 
    \label{mass_distributionb05}
\end{figure}

\begin{figure}    
    \includegraphics[width=95 mm]{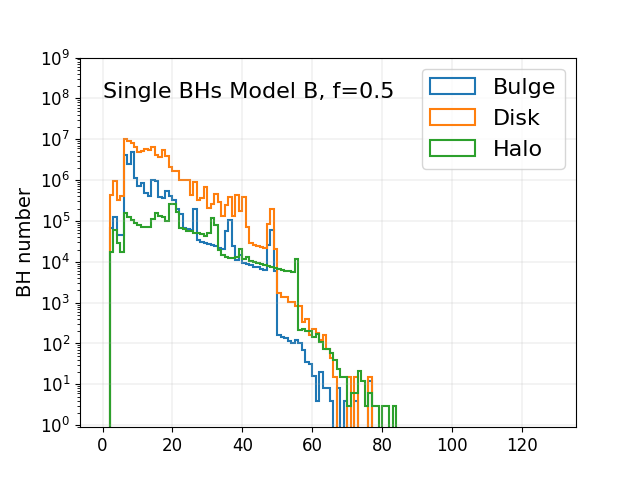} 
    \caption{Single BH mass distribution in three Galactic components for model B. In model B we assume that CE phase with HG star donor always leads to binary system merger. Mass of the objects formed in merger assuming 50 \% mass loss from the second merging object (except for two compact object mergers).} 
    \label{mass_distributionad05}
\end{figure}

\begin{figure} 

    \includegraphics[width=95mm]{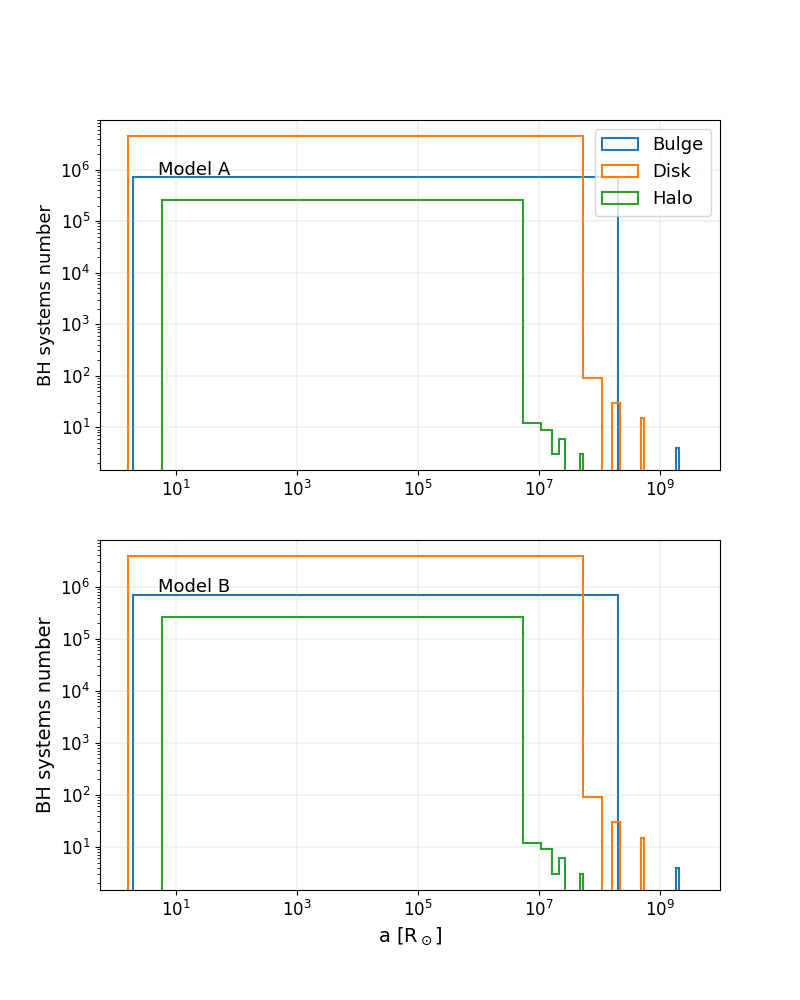} 
    \caption{Distribution of separation in all Galactic BH binary systems in three Galactic components (bulge, disk, halo) and for CE model A (top) and B (bottom).} 
    \label{separation_distribution}
\end{figure}

\begin{figure}    
    \includegraphics[width=95 mm]{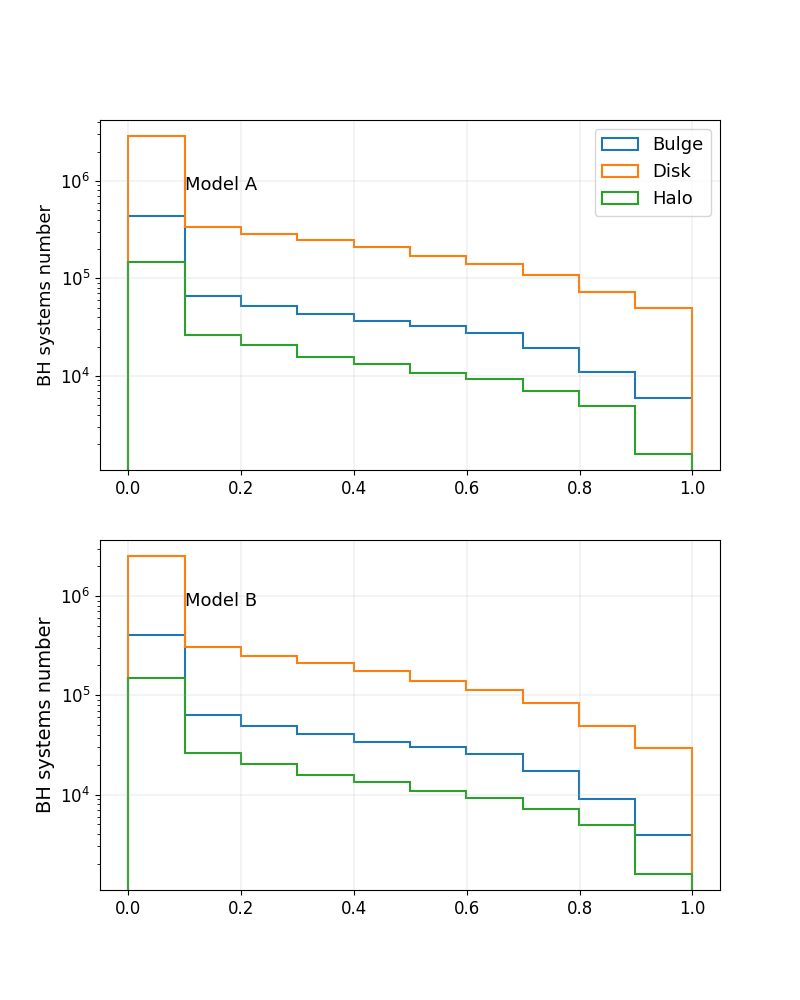} 
    \caption{Distribution of eccentricity in all Galactic BH binary systems in three Galactic components (bulge, disk, halo) and for CE model A (top) and B (bottom).} 
    \label{eccentricity_distribution}
\end{figure}

\begin{figure} 

    \includegraphics[width=95mm]{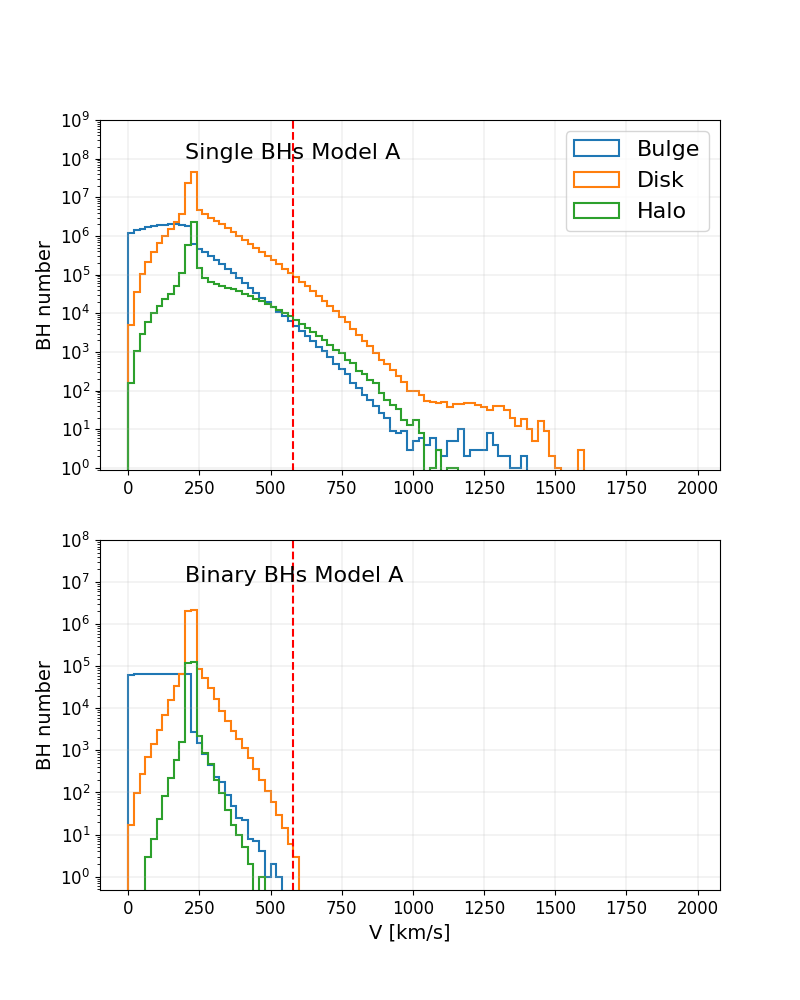} 
    \caption{Single (top) and binary (bottom) BHs velocity distribution in three Galactic components for model A. In model A we allow binary system to surrvive CE phase with HG donor star. With the red, dashed line we marked the velocity equal 580 km/s which is an estimated value of the local Galactic escape speed at the Sun's position \citep{2018A&A...616L...9M}. \color{black}} 
    \label{velocity_distributionb}
\end{figure}

\begin{figure}    
    \includegraphics[width=95mm]{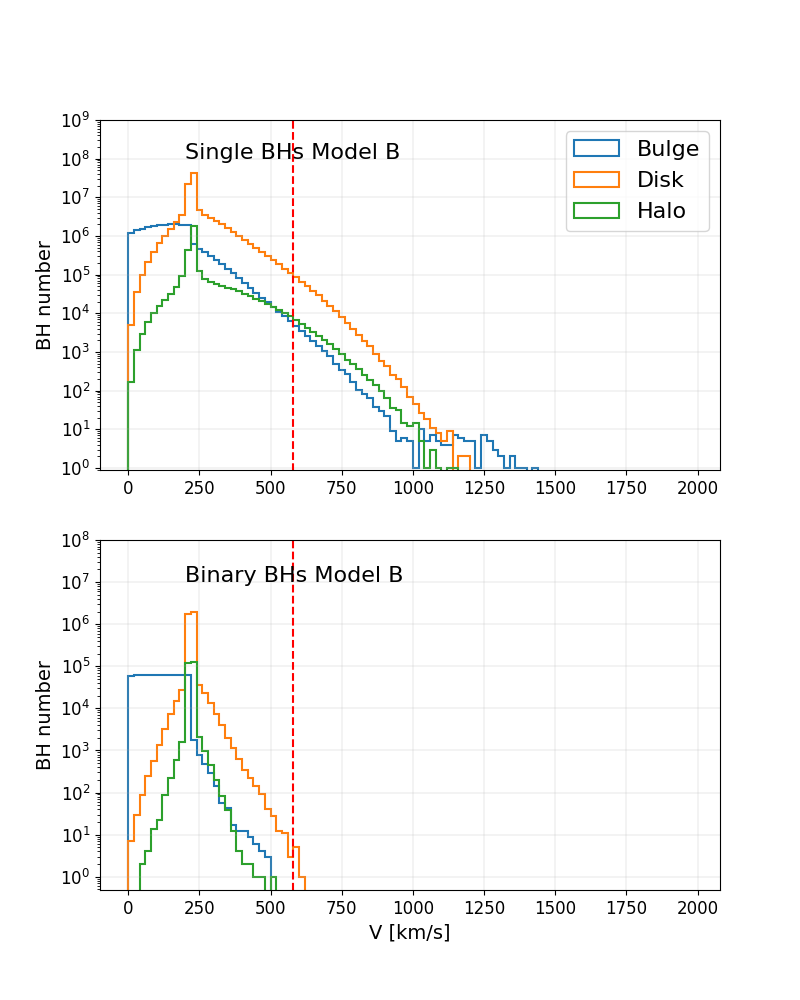}
    \caption{Single (top) and binary (bottom) BHs velocity distribution in three Galactic components for model B. In model B we assume that CE phase with HG donor star always leads to binary system merger. With the red, dashed line we marked the velocity equal 580 km/s which is an estimated value of the local Galactic escape speed at the Sun's position \citep{2018A&A...616L...9M}.} 
    \label{velocity_distributionad}
\end{figure}

\begin{figure}[hbt!] 
 \centering
 \includegraphics[width=90mm]{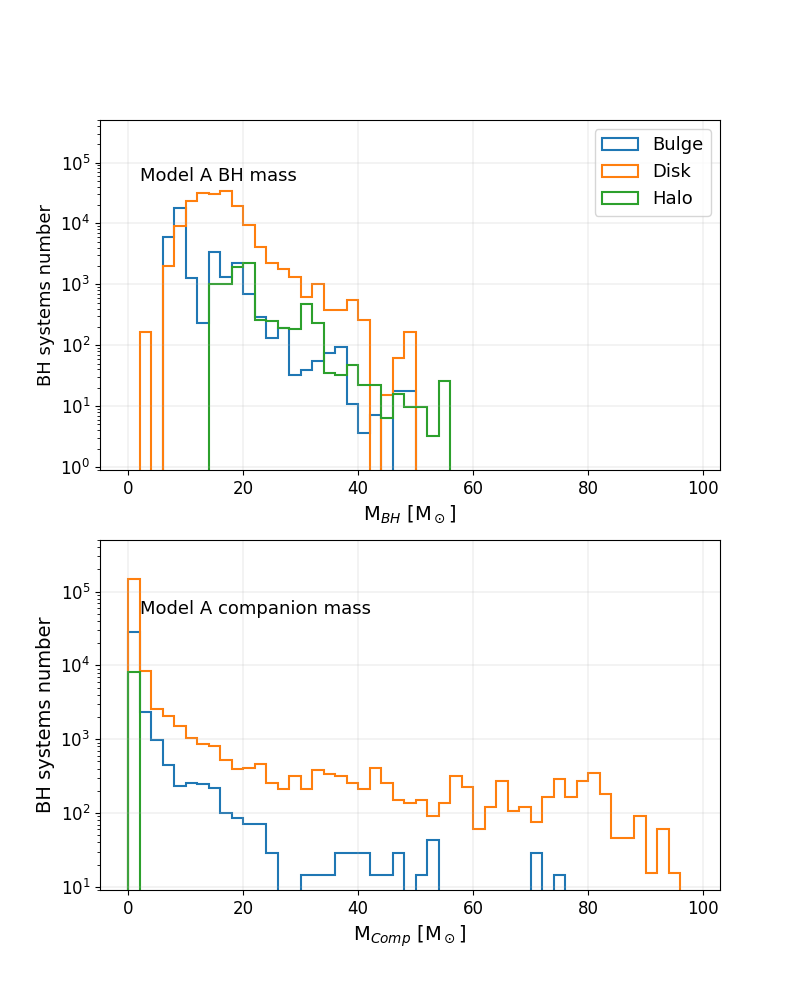}
 \caption{ Distribution of BH (top) and its companion masses (botton) in Galactic BH binary systems with non evolved companion in three Galactic components, for evolutionary model A.}
 \label{fig:BH_binaries_mass}
\end{figure}

\begin{figure}[hbt!] 
 \centering
 \includegraphics[width=90mm]{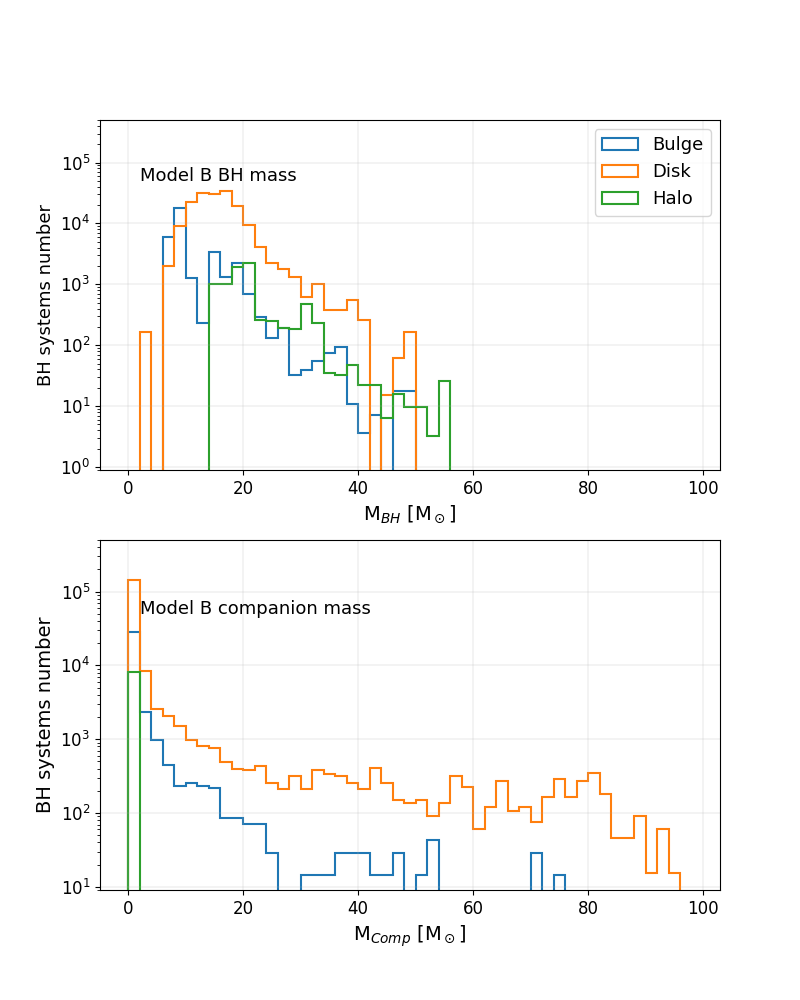}
 \caption{ Distribution of BH (top) and its companion masses (botton) in Galactic BH binary systems with non evolved companion in three Galactic components, for evolutionary model B.}
 \label{fig:BH_binaries_masB}
\end{figure}

\begin{figure}[hbt!] 
 \centering
 \includegraphics[width=90mm]{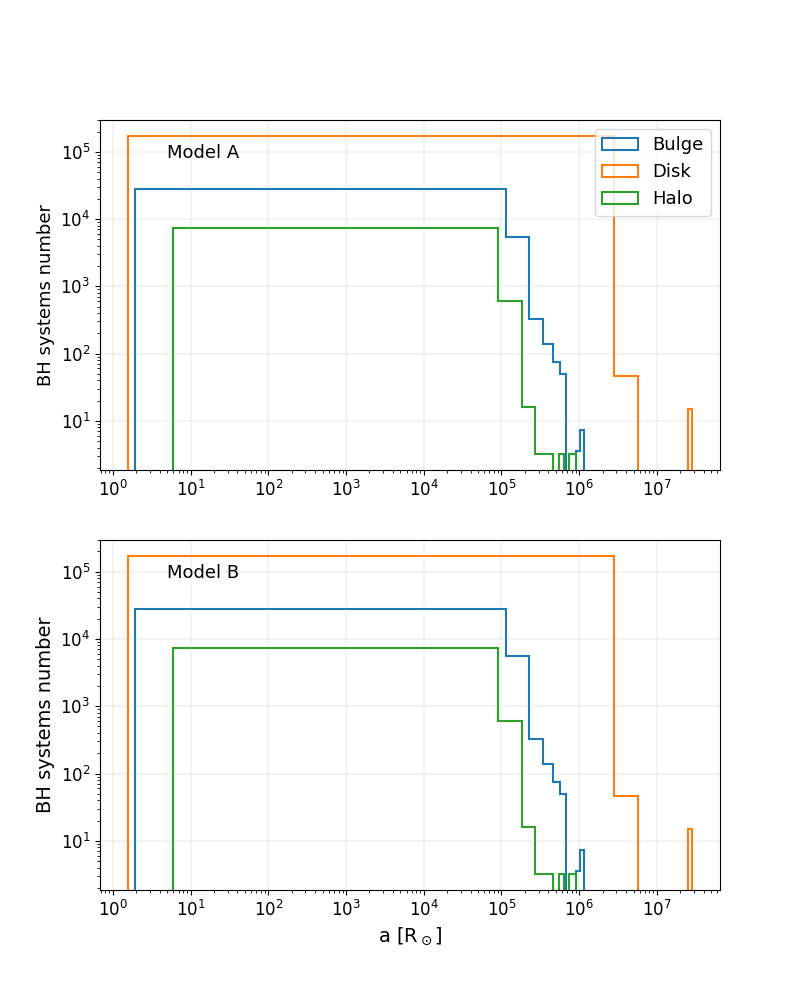}
 \caption{ Distribution of separation in Galactic BH binary systems with non evolved companion in three Galactic components, for evolutionary model A (top) and B (bottom).}
 \label{fig:BH_binaries_separation}
\end{figure}

\begin{figure}[hbt!] 
 \centering
 \includegraphics[width=90mm]{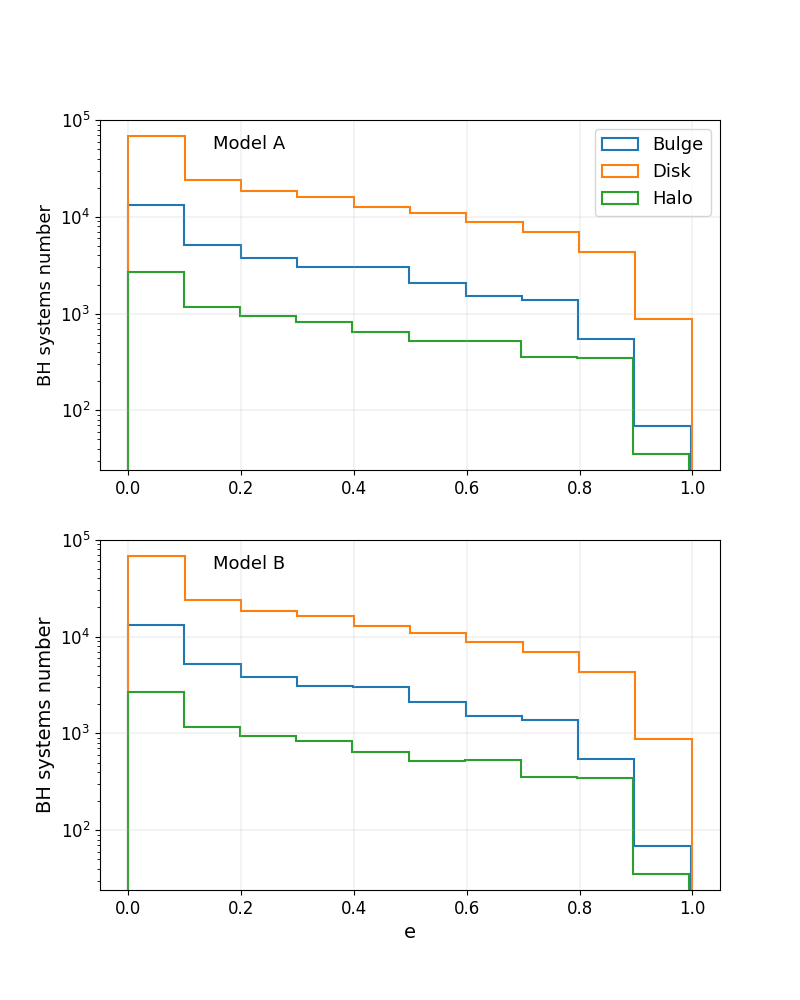}
 \caption{ Distribution of eccentricity in Galactic BH binary systems with non evolved companion in three Galactic components, for evolutionary model A (top) and B (bottom).}
 \label{fig:BH_binaries_eccentricity}
\end{figure}

\begin{figure}[hbt!] 
 \centering
 \includegraphics[width=90mm]{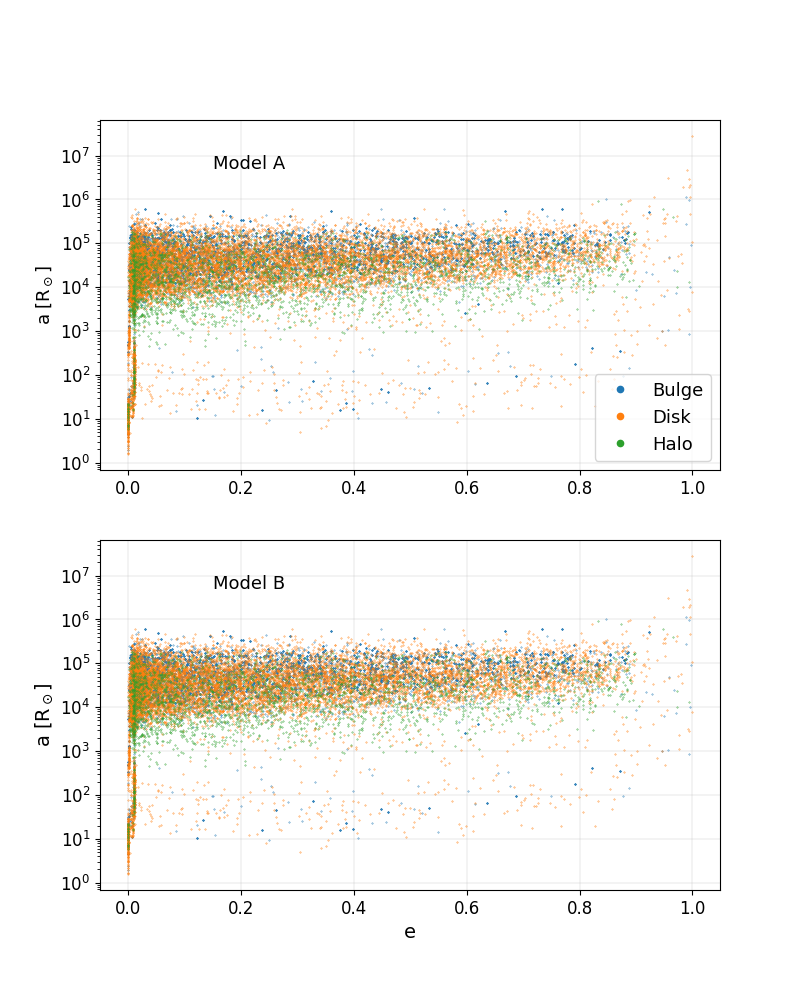}
 \caption{ Diagram with orbital parameters - separation and eccentricity for BH binary systems with non compact companion. Results for two models A (top) and B (bottom).}
 \label{fig:BH_binaries_e_a}
\end{figure}

We calculated how merger rates of double compact object systems (BH-BH, BH-NS and NS-NS) have been changing since Galactic formation till the current moment. We present results for two models, A and B (Sec. \ref{sec:models}). 
 DCO merger rates have been changing along with star formation and stellar metallicity of Galactic components (see Fig. \ref{fig:merger_rates}). In general the higher star formation rate in the given time, the higher merger rates, as the time-delay distribution between formation of binary system and its merger is a steep power-low ($\propto t^{-1}$)\citep{2012ApJ...759...52D}. Also the metallicty of stellar population is an important factor which strongly influences DCO merger rates, especially for BH-BH systems \citep{2013ApJ...779...72D,2015MNRAS.451.4086S,2016MNRAS.461.3877D}. 
The highest BH-BH merger rates in both models A and B occurred between 8-11 Gyr ago. The effect was caused by a peak in SFR  (Sec. \ref{sec:sfr}) as at that time there was an intensified star formation episode in bulge, the stellar population of thick disk was forming and star formation in thin disk have already started. High BH-BH merger rates were also caused by low metallicity of stars forming in this period. In both models BH-BH merger rates are going down with time but in model B rates decrease more dramatically. In higher metallicity, due to increased mass loss in stellar winds, binary systems evolve in a such way that HG star is more often a donor in common envelope phase. These systems are then eliminated from a binary system population (see Sec. \ref{sec:models}). \\
Merger rates of BH-NS and NS-NS systems did not change so much with Galaxy formation and metallicity as rates for BH-BH systems. Lower metallicity slightly lowers NS-NS merger rates but this effect was compensated by more intensive SFR episodes in early ages (8-11 Gyr). 
Because in standard physical model we assumed natal kicks with high $\sigma$=265 km/s (see Sec. \ref{sec:physicamodell}) and fallback factor inversely proportional to mass, many low mass compact object binary systems get disrupted by SN explosion/core collapse. Those disruptions decreased compact object merger rates, especially for NS-NS systems. Our rates for double compact object systems mergers can be compared with the results for other implemented physical models for wide range of natal kick distributions, common envelope efficiency parameters or mass fraction ejected during RLOF which are presented in for example: \cite{1992ApJ...386..197T, 2010CQGra..27q3001A, 10.1046/j.1365-8711.2003.06616.x, 2018arXiv181210065B}. \\
For comparison we present merger rates evolution and current values for old, simplified Galactic SFR and chemical evolution model (Fig. \ref{fig:merger_rates2}). The previous model assumed only one Galactic component (disk) with mass $3.5 \times 10^{10}$ M$\odot$. The SFR was constant during 10 Gyrs (3.5 M$_\odot/$year) and all stars were formed with same metallicity equal Z$_\odot$=0.014.  \\
 In the Tab. \ref{tab:mergerates} we present current Galactic merger rates of BH-BH, BH-NS and NS-NS systems per {Myr$^{-1}$} for two considered CE models A and B. For comparison we also calculated current merger rates for old SFR and metallicity distribution model of Milky Way (Sec. \ref{sec:sfr}). Merger rates for the old model are in general lower than for the new model for all types of DCO systems. The most significant difference is in the case of BH-BH systems for which current rates (in model B) are one order of magnitude lower for the old model.    
    
\color{black}

   \begin{table} 
      \caption[]{Current Galactic merger rates for new and old Galactic SFR and chemical evolution models. Results for evolutionary model A and B are given in order A/(B) (see Sec. \ref{sec:models})} \label{tab:mergerates}
     $$ 
         \begin{array}{lrr}
            \hline
            \hline
            \noalign{\smallskip}
            \mbox{System type} & \mbox{New SFR model} & \mbox{Old SFR model} \\
            {}&\mbox{[Myr}{^{-1}]}&\mbox{[Myr}{^{-1}]}\\
            \noalign{\smallskip}
            \hline
            \hline
            \noalign{\smallskip}
            \mbox{BH-BH} & 81.1 /(3.1) & 36.2/(0.3)\\
            \mbox{BH-NS} & 8.5 /(0.7) & 2.3/(0.2) \\
            \mbox{NS-NS} & 59.0  /(14.1) & 46.9/(12.9)\\
            \hline
            \hline
         \end{array}
     $$ 
   \end{table}
   
   We calculated the fractions of current Galactic double compact object systems that will merge in Hubble time (in next 14 Gyr). Percentage fraction such systems for two considered evolution models given in order A/(B) is: $\sim$6\%/(1\%) for BH-BH, $\sim$13\%/(6\%) for NS-NS and $\sim$16\%/(2\%) for BH-NS systems. Note that star formation is still taking place in some parts of the Milky Way so number of double compact object systems will also increase. 
   Numeric data are presented in Table \ref{tab:mergersystems}:
   
      \begin{table} 
      \caption[]{Number of current Galactic double compact systems that will merge in time shorter/longer than Hubble time (T$_{Hub}$). Results for model A and B, see section \ref{sec:models} and new SFR and metallicity distribution model\ref{sec:sfr} } \label{tab:mergersystems}
     $$ 
         \begin{array}{lrr}
            \hline
            \hline
            \noalign{\smallskip}
            \mbox{System type} & \mbox{Merger time} < \mbox{T}_{Hub} & \mbox{Merger time} > \mbox{T}_{Hub}   \\
            \noalign{\smallskip}
            \hline
            \hline
            \noalign{\smallskip}
            \mbox{BH-BH} & 1.8 \times 10^5 /(1.7 \times 10^4) & 3.4 \times 10^6/(2.5 \times 10^6) \\
            \mbox{NS-NS} & 6.1 \times 10^4 /(2.9 \times 10^4)  & 4.7 \times 10^5/(3.5 \times 10^5)\\
            \mbox{BH-NS} & 3.9 \times 10^4 /(2.9 \times 10^3) & 2.4 \times 10^5/(1.3 \times 10^5) \\
            \hline
            \hline
         \end{array}
     $$ 
   \end{table}

We do not compare our current Galactic results with LIGO/Virgo estimates since to calculate cosmic merger rate of DCO one needs to take into account entire cosmic star formation rates SFR(z) and metallicity distribution Z(z) as a function of redshift. However, such calculation for StarTrack physical models (including presented in this work) were carried out in \cite{belczynski2017evolutionary}, Sec. 3.2 noting full agreement with BH-BH, BH-NS and NS-NS LIGO/Virgo estimates.

\section{Conclusions}

We present population synthesis statistical estimates of the current Milky Way black hole population properties. We used the most current version of the StarTrack code with standard physics (Sec. \ref{sec:method}) and processed the data with the new star formation rates and metallicity distribution model of our Galaxy, based on theoretical models and observations. We show results for two models: A and B, which correspond to different scenarios of a CE phase (Sec. \ref{sec:models}).  We find that: \\
1) At the current moment the Milky Way (disk+bulge+halo) contains about $1.2 \times 10^8$ single black holes with average mass 14 M$_\odot$ and $9.3 \times 10^6$ black holes in binary systems with average mass 19 M$_\odot$. \\
2) There are three main formation channels of single BHs: $\sim 50 \%$ are remnants of massive single star evolution, $\sim 30\%$ formed in binary systems merger, $\sim 20\%$ of single BHs originate from disrupted binary systems during black hole/neutron  star formation.\\
3) The most massive black hole in simulation comes from old, low in metal environment of Galactic halo. It formed in BH-MS system coalescence and its mass is as large as $\sim$ 113 M$_\odot$. \\
4) Black holes in binary systems constitute $\sim$ 10 \% of the whole Galactic BH population. Most of BHs in binary systems are in BH-BH configuration. The fraction of black hole binaries with non compact companion is small, about 0.3 \% of all Galactic BHs. \\
-We estimate how double compact object systems merger rates (BH-BH, BH-NS and NS-NS) have changed along with the Galaxy star formation. Current Galactic merger rates depend on model and they are estimated at $\sim$ 81/3 Myr$^{-1}$ for BH-BH, $\sim$ 9/1 Myr$^{-1}$ for BH-NS and $\sim$ 59/14 Myr$^{-1}$ for NS-NS systems.  \\
5) We constrain that only $\sim$ 0.006 \% of total Galactic halo mass (including dark matter) could be hidden in the form of stellar origin BHs which are not detectable by current
observational surveys \\
6) Only $\sim$ 5 \% of single BHs and 0.001 \% of binary BHs have enough high velocities to escape from Galactic potential. 

It is worth to mention that due to our assumption on binary fraction Sec. (\ref{sec:physicamodell}) and orbital seperation (Sec. \ref{sec:initaial}) number of binary systems as well as system mergers could be slightly underestimated.

\clearpage

\clearpage

\clearpage

\clearpage

\clearpage

\clearpage

   \begin{table*} 
      \caption[]{Types of objects originated from different types of stellar mergers based on Table 2 from \cite{Hurley_2002}.}
        \label{tab:merger_types}
     $$ 
         \begin{array}{ccccccccccccccccc}
            \hline
            \noalign{\smallskip}
            \multicolumn{10}{c}{\mbox{First star type k$_1$}}    \\
            \noalign{\smallskip}
            &  & 0 & 1 & 2 & 3 & 4 & 5 & 6 & 7 & 8 & 9 & 10 & 11 & 12 & 13 & 14\\
            \noalign{\smallskip}
            \hline
            \noalign{\smallskip}
             & 0 & 1 & 1 & 2 & 3 & 4 & 5 & 6 & 4 & 6 & 6 & 3 & 6 & 6 & 13 & 14\\
             & 1 & 1 & 1 & 2 & 3 & 4 & 5 & 6 & 4 & 6 & 6 & 3 & 6 & 6 & 13 & 14\\
             & 2 & 2 & 2 & 3 & 3 & 4 & 4 & 5 & 4 & 4 & 4 & 3 & 5 & 5 & 13 & 14\\
             & 3 & 3 & 3 & 3 & 3 & 4 & 4 & 5 & 4 & 4 & 4 & 3 & 5 & 5 & 13 & 14\\
             & 4 & 4 & 4 & 4 & 4 & 4 & 4 & 4 & 4 & 4 & 4 & 4 & 4 & 4 & 13 & 14\\
         \mbox{Second} & 5 & 5 & 5 & 4 & 4 & 4 & 4 & 4 & 4 & 4 & 4 & 4 & 4 & 4 & 13 & 14\\
          \mbox{star} & 6 & 6 & 6 & 5 & 5 & 4 & 4 & 6 & 4 & 6 & 6 & 5 & 6 & 6 & 13 & 14\\
          \mbox{type} & 7 & 4 & 4 & 4 & 4 & 4 & 4 & 4 & 7 & 8 & 9 & 7 & 9 & 9 & 13 & 14\\
         \mbox{k}_2 & 8 & 6 & 6 & 4 & 4 & 4 & 4 & 6 & 8 & 8 & 9 & 7 & 9 & 9 & 13 & 14\\
             & 9 & 6 & 6 & 4 & 4 & 4 & 4 & 6 & 9 & 9 & 9 & 7 & 9 & 9 & 13 & 14\\
             & 10 & 3 & 3 & 3 & 3 & 4 & 4 & 5 & 7 & 7 & 7 & 15 & 9 & 9 & 13 & 14\\
             & 11 & 6 & 6 & 5 & 5 & 4 & 4 & 6 & 9 & 9 & 9 & 9 & 11 & 12 & 13 & 14\\
             & 12 & 6 & 6 & 5 & 5 & 4 & 4 & 6 & 9 & 9 & 9 & 9 & 12 & 12 & 13 & 14 \\
             & 13 & 13 & 13 & 13 & 13 & 13 & 13 & 13 & 13 & 13 & 13 & 13 & 13 & 13 & 13 & 14\\
             & 14 & 14 & 14 & 14 & 14 & 14 & 14 & 14 & 14 & 14 & 14 & 14 & 14 & 14 & 14 & 14\\
            \hline
         \end{array}
     $$ 
    \\
    
The numeric types are consistent with \cite{Hurley_2002}: \\
0 – main sequence star with M<=0.7 Msun (deeply or fully convective) \\
1 – main sequence star with M>0.7 Msun \\
2 – Hertzsprung gap star \\
3 – first giant branch star \\
4 – core helium burning star \\
5 – early asymptotic giant branch star \\
6 – thermally pulsing asymptotic giant branch star \\
7 – main sequence naked helium star \\
8 – Hertzsprung gap naked helium star \\
9 – giant branch naked helium star \\
10 – helium white dwarf \\
11 – carbon/oxygen white dwarf  \\
12 – oxygen/neon white dwarf \\
13 – neutron star \\
14 – black hole \\
15 – massless remnant \\
   \end{table*}

\clearpage

\begin{acknowledgements}
      Authors acknowledge support from the Polish National \\ Science Center (NCN) grant:  project Maestro 2018/30/A/ST9/00050 \\
      We would like to thank: Grzegorz Wiktorowicz, Pawel Pietrukowicz and Thomas Bensby for their comments and advices.
\end{acknowledgements}

\bibliographystyle{aa}
\bibliography{ms}

\begin{thebibliography}{107}
\expandafter\ifx\csname natexlab\endcsname\relax\def\natexlab#1{#1}\fi

\bibitem[{{Abadie} {et~al.}(2010){Abadie}, {Abbott}, {Abbott}, {Abernathy},
  {Accadia}, {Acernese}, {Adams}, {Adhikari}, {Ajith}, {Allen}, {Allen},
  {Amador Ceron}, {Amin}, {Anderson}, {Anderson}, {Antonucci}, {Aoudia},
  {Arain}, {Araya}, {Aronsson}, {Arun}, {Aso}, {Aston}, {Astone}, {Atkinson},
  {Aufmuth}, {Aulbert}, {Babak}, {Baker}, {Ballardin}, {Ballmer}, {Barker},
  {Barnum}, {Barone}, {Barr}, {Barriga}, {Barsotti}, {Barsuglia}, {Barton},
  {Bartos}, {Bassiri}, {Bastarrika}, {Bauchrowitz}, {Bauer}, {Behnke}, {Beker},
  {Belczynski}, {Benacquista}, {Bertolini}, {Betzwieser}, {Beveridge},
  {Beyersdorf}, {Bigotta}, {Bilenko}, {Billingsley}, {Birch}, {Birindelli},
  {Biswas}, {Bitossi}, {Bizouard}, {Black}, {Blackburn}, {Blackburn}, {Blair},
  {Bland}, {Blom}, {Blomberg}, {Boccara}, {Bock}, {Bodiya}, {Bondarescu},
  {Bondu}, {Bonelli}, {Bork}, {Born}, {Bose}, {Bosi}, {Boyle}, {Braccini},
  {Bradaschia}, {Brady}, {Braginsky}, {Brau}, {Breyer}, {Bridges}, {Brillet},
  {Brinkmann}, {Brisson}, {Britzger}, {Brooks}, {Brown}, {Budzy{\'n}ski},
  {Bulik}, {Bulten}, {Buonanno}, {Burguet-Castell}, {Burmeister}, {Buskulic},
  {Byer}, {Cadonati}, {Cagnoli}, {Calloni}, {Camp}, {Campagna}, {Campsie},
  {Cannizzo}, {Cannon}, {Canuel}, {Cao}, {Capano}, {Carbognani}, {Caride},
  {Caudill}, {Cavagli{\`a}}, {Cavalier}, {Cavalieri}, {Cella}, {Cepeda},
  {Cesarini}, {Chalermsongsak}, {Chalkley}, {Charlton}, {Chassand e Mottin},
  {Chelkowski}, {Chen}, {Chincarini}, {Christensen}, {Chua}, {Chung}, {Clark},
  {Clark}, {Clayton}, {Cleva}, {Coccia}, {Colacino}, {Colas}, {Colla},
  {Colombini}, {Conte}, {Cook}, {Corbitt}, {Corda}, {Cornish}, {Corsi},
  {Costa}, {Coulon}, {Coward}, {Coyne}, {Creighton}, {Creighton}, {Cruise},
  {Culter}, {Cumming}, {Cunningham}, {Cuoco}, {Dahl}, {Danilishin},
  {Dannenberg}, {D'Antonio}, {Danzmann}, {Dari}, {Das}, {Dattilo}, {Daudert},
  {Davier}, {Davies}, {Davis}, {Daw}, {Day}, {Dayanga}, {De Rosa}, {DeBra},
  {Degallaix}, {del Prete}, {Dergachev}, {DeRosa}, {DeSalvo}, {Devanka},
  {Dhurandhar}, {Di Fiore}, {Di Lieto}, {Di Palma}, {Emilio}, {Di Virgilio},
  {D{\'\i}az}, {Dietz}, {Donovan}, {Dooley}, {Doomes}, {Dorsher}, {Douglas},
  {Drago}, {Drever}, {Driggers}, {Dueck}, {Dumas}, {Eberle}, {Edgar},
  {Edwards}, {Effler}, {Ehrens}, {Engel}, {Etzel}, {Evans}, {Evans}, {Fafone},
  {Fairhurst}, {Fan}, {Farr}, {Fazi}, {Fehrmann}, {Feldbaum}, {Ferrante},
  {Fidecaro}, {Finn}, {Fiori}, {Flaminio}, {Flanigan}, {Flasch}, {Foley},
  {Forrest}, {Forsi}, {Fotopoulos}, {Fournier}, {Franc}, {Frasca}, {Frasconi},
  {Frede}, {Frei}, {Frei}, {Freise}, {Frey}, {Fricke}, {Friedrich},
  {Fritschel}, {Frolov}, {Fulda}, {Fyffe}, {Gammaitoni}, {Garofoli}, {Garufi},
  {Gemme}, {Genin}, {Gennai}, {Gholami}, {Ghosh}, {Giaime}, {Giampanis},
  {Giardina}, {Giazotto}, {Gill}, {Goetz}, {Goggin}, {Gonz{\'a}lez},
  {Gorodetsky}, {Go{\ss}ler}, {Gouaty}, {Graef}, {Granata}, {Grant}, {Gras},
  {Gray}, {Greenhalgh}, {Gretarsson}, {Greverie}, {Grosso}, {Grote},
  {Grunewald}, {Guidi}, {Gustafson}, {Gustafson}, {Hage}, {Hall}, {Hallam},
  {Hammer}, {Hammond}, {Hanks}, {Hanna}, {Hanson}, {Harms}, {Harry}, {Harry},
  {Harstad}, {Haughian}, {Hayama}, {Heefner}, {Heitmann}, {Hello}, {Heng},
  {Heptonstall}, {Hewitson}, {Hild}, {Hirose}, {Hoak}, {Hodge}, {Holt},
  {Hosken}, {Hough}, {Howell}, {Hoyland}, {Huet}, {Hughey}, {Husa}, {Huttner},
  {Huynh-Dinh}, {Ingram}, {Inta}, {Isogai}, {Ivanov}, {Jaranowski}, {Johnson},
  {Jones}, {Jones}, {Jones}, {Ju}, {Kalmus}, {Kalogera}, {Kandhasamy},
  {Kanner}, {Katsavounidis}, {Kawabe}, {Kawamura}, {Kawazoe}, {Kells},
  {Keppel}, {Khalaidovski}, {Khalili}, {Khazanov}, {Kim}, {Kim}, {King},
  {Kinzel}, {Kissel}, {Klimenko}, {Kondrashov}, {Kopparapu}, {Koranda},
  {Kowalska}, {Kozak}, {Krause}, {Kringel}, {Krishnamurthy}, {Krishnan},
  {Kr{\'o}lak}, {Kuehn}, {Kullman}, {Kumar}, {Kwee}, {Landry}, {Lang}, {Lantz},
  {Lastzka}, {Lazzarini}, {Leaci}, {Leong}, {Leonor}, {Leroy}, {Letendre},
  {Li}, {Li}, {Lin}, {Lindquist}, {Lockerbie}, {Lodhia}, {Lorenzini},
  {Loriette}, {Lormand}, {Losurdo}, {Lu}, {Luan}, {Lubi{\'n}ski}, {Lucianetti},
  {L{\"u}ck}, {Lundgren}, {Machenschalk}, {MacInnis}, {Mackowski},
  {Mageswaran}, {Mailand }, {Majorana}, {Mak}, {Man}, {Mandel}, {Mandic},
  {Mantovani}, {Marchesoni}, {Marion}, {M{\'a}rka}, {M{\'a}rka}, {Maros},
  {Marque}, {Martelli}, {Martin}, {Martin}, {Marx}, {Mason}, {Masserot},
  {Matichard}, {Matone}, {Matzner}, {Mavalvala}, {McCarthy}, {McClelland},
  {McGuire}, {McIntyre}, {McIvor}, {McKechan}, {Meadors}, {Mehmet}, {Meier},
  {Melatos}, {Melissinos}, {Mendell}, {Men{\'e}ndez}, {Mercer}, {Merill},
  {Meshkov}, {Messenger}, {Meyer}, {Miao}, {Michel}, {Milano}, {Miller},
  {Minenkov}, {Mino}, {Mitra}, {Mitrofanov}, {Mitselmakher}, {Mittleman},
  {Moe}, {Mohan}, {Mohanty}, {Mohapatra}, {Moraru}, {Moreau}, {Moreno},
  {Morgado}, {Morgia}, {Morioka}, {Mors}, {Mosca}, {Moscatelli}, {Mossavi},
  {Mours}, {MowLowry}, {Mueller}, {Mukherjee}, {Mullavey},
  {M{\"u}ller-Ebhardt}, {Munch}, {Murray}, {Nash}, {Nawrodt}, {Nelson}, {Neri},
  {Newton}, {Nishizawa}, {Nocera}, {Nolting}, {Ochsner}, {O'Dell}, {Ogin},
  {Oldenburg}, {O'Reilly}, {O'Shaughnessy}, {Osthelder}, {Ottaway}, {Ottens},
  {Overmier}, {Owen}, {Page}, {Pagliaroli}, {Palladino}, {Palomba}, {Pan},
  {Pankow}, {Paoletti}, {Papa}, {Pardi}, {Pareja}, {Parisi}, {Pasqualetti},
  {Passaquieti}, {Passuello}, {Patel}, {Pedraza}, {Pekowsky}, {Penn},
  {Peralta}, {Perreca}, {Persichetti}, {Pichot}, {Pickenpack}, {Piergiovanni},
  {Pietka}, {Pinard}, {Pinto}, {Pitkin}, {Pletsch}, {Plissi}, {Poggiani},
  {Postiglione}, {Prato}, {Predoi}, {Price}, {Prijatelj}, {Principe},
  {Privitera}, {Prix}, {Prodi}, {Prokhorov}, {Puncken}, {Punturo}, {Puppo},
  {Quetschke}, {Raab}, {Rabaste}, {Rabeling}, {Radke}, {Radkins}, {Raffai},
  {Rakhmanov}, {Rankins}, {Rapagnani}, {Raymond}, {Re}, {Reed}, {Reed},
  {Regimbau}, {Reid}, {Reitze}, {Ricci}, {Riesen}, {Riles}, {Roberts},
  {Robertson}, {Robinet}, {Robinson}, {Robinson}, {Rocchi}, {Roddy},
  {R{\"o}ver}, {Rogstad}, {Rolland}, {Rollins}, {Romano}, {Romano}, {Romie},
  {Rosi{\'n}ska}, {Rowan}, {R{\"u}diger}, {Ruggi}, {Ryan}, {Sakata}, {Sakosky},
  {Salemi}, {Sammut}, {Sancho de la Jordana}, {Sandberg}, {Sannibale},
  {Santamar{\'\i}a}, {Santostasi}, {Saraf}, {Sassolas}, {Sathyaprakash},
  {Sato}, {Satterthwaite}, {Saulson}, {Savage}, {Schilling}, {Schnabel},
  {Schofield}, {Schulz}, {Schutz}, {Schwinberg}, {Scott}, {Scott}, {Searle},
  {Seifert}, {Sellers}, {Sengupta}, {Sentenac}, {Sergeev}, {Shaddock},
  {Shapiro}, {Shawhan}, {Shoemaker}, {Sibley}, {Siemens}, {Sigg}, {Singer},
  {Sintes}, {Skelton}, {Slagmolen}, {Slutsky}, {Smith}, {Smith}, {Smith},
  {Somiya}, {Sorazu}, {Speirits}, {Stein}, {Stein}, {Steinlechner},
  {Steplewski}, {Stochino}, {Stone}, {Strain}, {Strigin}, {Stroeer}, {Sturani},
  {Stuver}, {Summerscales}, {Sung}, {Susmithan}, {Sutton}, {Swinkels},
  {Talukder}, {Tanner}, {Tarabrin}, {Taylor}, {Taylor}, {Thomas}, {Thorne},
  {Thorne}, {Thrane}, {Th{\"u}ring}, {Titsler}, {Tokmakov}, {Toncelli},
  {Tonelli}, {Torres}, {Torrie}, {Tournefier}, {Travasso}, {Traylor}, {Trias},
  {Trummer}, {Tseng}, {Ugolini}, {Urbanek}, {Vahlbruch}, {Vaishnav}, {Vajente},
  {Vallisneri}, {van den Brand}, {Van Den Broeck}, {van der Putten}, {van der
  Sluys}, {van Veggel}, {Vass}, {Vaulin}, {Vavoulidis}, {Vecchio}, {Vedovato},
  {Veitch}, {Veitch}, {Veltkamp}, {Verkindt}, {Vetrano}, {Vicer{\'e}},
  {Villar}, {Vinet}, {Vocca}, {Vorvick}, {Vyachanin}, {Waldman}, {Wallace},
  {Wanner}, {Ward}, {Was}, {Wei}, {Weinert}, {Weinstein}, {Weiss}, {Wen},
  {Wen}, {Wessels}, {West}, {Westphal}, {Wette}, {Whelan}, {Whitcomb}, {White},
  {Whiting}, {Wilkinson}, {Willems}, {Williams}, {Willke}, {Winkelmann},
  {Winkler}, {Wipf}, {Wiseman}, {Woan}, {Wooley}, {Worden}, {Yakushin},
  {Yamamoto}, {Yamamoto}, {Yeaton-Massey}, {Yoshida}, {Yu}, {Yvert}, {Zanolin},
  {Zhang}, {Zhang}, {Zhao}, {Zotov}, {Zucker}, {Zweizig}, {LIGO Scientific
  Collaboration}, \& {Virgo Collaboration}}]{2010CQGra..27q3001A}
{Abadie}, J., {Abbott}, B.~P., {Abbott}, R., {et~al.} 2010, Classical and
  Quantum Gravity, 27, 173001

\bibitem[{{Abbott} {et~al.}(2016{\natexlab{a}}){Abbott}, {Abbott}, {Abbott},
  {Abernathy}, {Acernese}, {Ackley}, {Adams}, {Adams}, {Addesso}, \&
  {Adhikari}}]{2016PhRvX...6d1015A}
{Abbott}, B.~P., {Abbott}, R., {Abbott}, T.~D., {et~al.} 2016{\natexlab{a}},
  Physical Review X, 6, 041015

\bibitem[{{Abbott} {et~al.}(2016{\natexlab{b}}){Abbott}, {Abbott}, {Abbott},
  {Abernathy}, {Acernese}, {Ackley}, {Adams}, {Adams}, {Addesso}, \&
  {Adhikari}}]{2016PhRvL.116x1103A}
{Abbott}, B.~P., {Abbott}, R., {Abbott}, T.~D., {et~al.} 2016{\natexlab{b}},
  \prl, 116, 241103

\bibitem[{{Abbott} {et~al.}(2016{\natexlab{c}}){Abbott}, {Abbott}, {Abbott},
  {Abernathy}, {Acernese}, {Ackley}, {Adams}, {Adams}, {Addesso}, \&
  {Adhikari}}]{2016PhRvL.116f1102A}
{Abbott}, B.~P., {Abbott}, R., {Abbott}, T.~D., {et~al.} 2016{\natexlab{c}},
  \prl, 116, 061102

\bibitem[{{Abbott} {et~al.}(2017{\natexlab{a}}){Abbott}, {Abbott}, {Abbott},
  {Acernese}, {Ackley}, {Adams}, {Adams}, {Addesso}, {Adhikari}, \&
  {Adya}}]{2017PhRvL.118v1101A}
{Abbott}, B.~P., {Abbott}, R., {Abbott}, T.~D., {et~al.} 2017{\natexlab{a}},
  \prl, 118, 221101

\bibitem[{{Abbott} {et~al.}(2017{\natexlab{b}}){Abbott}, {Abbott}, {Abbott},
  {Acernese}, {Ackley}, {Adams}, {Adams}, {Addesso}, {Adhikari}, \&
  {Adya}}]{2017ApJ...851L..35A}
{Abbott}, B.~P., {Abbott}, R., {Abbott}, T.~D., {et~al.} 2017{\natexlab{b}},
  \apjl, 851, L35

\bibitem[{{Abbott} {et~al.}(2017{\natexlab{c}}){Abbott}, {Abbott}, {Abbott},
  {Acernese}, {Ackley}, {Adams}, {Adams}, {Addesso}, {Adhikari}, \&
  {Adya}}]{2017PhRvL.119n1101A}
{Abbott}, B.~P., {Abbott}, R., {Abbott}, T.~D., {et~al.} 2017{\natexlab{c}},
  \prl, 119, 141101

\bibitem[{{Abramowicz} {et~al.}(2018){Abramowicz}, {Bejger}, \&
  {Wielgus}}]{2018ApJ...868...17A}
{Abramowicz}, M.~A., {Bejger}, M., \& {Wielgus}, M. 2018, \apj, 868, 17

\bibitem[{{Alcock} {et~al.}(2001){Alcock}, {Allsman}, {Alves}, {Axelrod},
  {Becker}, {Bennett}, {Cook}, {Drake}, {Freeman}, {Geha}, {Griest}, {Lehner},
  {Marshall}, {Minniti}, {Nelson}, {Peterson}, {Popowski}, {Pratt}, {Quinn},
  {Stubbs}, {Sutherland}, {Tomaney}, {Vand ehei}, \&
  {Welch}}]{2001ApJS..136..439A}
{Alcock}, C., {Allsman}, R.~A., {Alves}, D.~R., {et~al.} 2001, \apjs, 136, 439

\bibitem[{{Antoniadis} {et~al.}(2013){Antoniadis}, {Freire}, {Wex}, {Tauris},
  {Lynch}, {van Kerkwijk}, {Kramer}, {Bassa}, {Dhillon}, {Driebe}, {Hessels},
  {Kaspi}, {Kondratiev}, {Langer}, {Marsh}, {McLaughlin}, {Pennucci}, {Ransom},
  {Stairs}, {van Leeuwen}, {Verbiest}, \& {Whelan}}]{2013Sci...340..448A}
{Antoniadis}, J., {Freire}, P.~C.~C., {Wex}, N., {et~al.} 2013, Science, 340,
  448

\bibitem[{{Antonini} {et~al.}(2017){Antonini}, {Toonen}, \&
  {Hamers}}]{2017ApJ...841...77A}
{Antonini}, F., {Toonen}, S., \& {Hamers}, A.~S. 2017, \apj, 841, 77

\bibitem[{{Arca-Sedda} {et~al.}(2018){Arca-Sedda}, {Li}, \&
  {Kocsis}}]{2018arXiv180506458A}
{Arca-Sedda}, M., {Li}, G., \& {Kocsis}, B. 2018, arXiv e-prints,
  arXiv:1805.06458

\bibitem[{{Asplund} {et~al.}(2009){Asplund}, {Grevesse}, {Sauval}, \&
  {Scott}}]{2009ARA&A..47..481A}
{Asplund}, M., {Grevesse}, N., {Sauval}, A.~J., \& {Scott}, P. 2009, \araa, 47,
  481

\bibitem[{{Belczynski} {et~al.}(2010){Belczynski}, {Bulik}, {Fryer}, {Ruiter},
  {Valsecchi}, {Vink}, \& {Hurley}}]{2010ApJ...714.1217B}
{Belczynski}, K., {Bulik}, T., {Fryer}, C.~L., {et~al.} 2010, \apj, 714, 1217

\bibitem[{{Belczynski} {et~al.}(2018){Belczynski}, {Bulik}, {Olejak},
  {Chruslinska}, {Singh}, {Pol}, {Zdunik}, {O'Shaughnessy}, {McLaughlin},
  {Lorimer}, {Korobkin}, {van den Heuvel}, {Davies}, \&
  {Holz}}]{2018arXiv181210065B}
{Belczynski}, K., {Bulik}, T., {Olejak}, A., {et~al.} 2018, arXiv e-prints,
  arXiv:1812.10065

\bibitem[{{Belczynski} {et~al.}(2016){Belczynski}, {Heger}, {Gladysz},
  {Ruiter}, {Woosley}, {Wiktorowicz}, {Chen}, {Bulik}, {O'Shaughnessy}, {Holz},
  {Fryer}, \& {Berti}}]{2016A&A...594A..97B}
{Belczynski}, K., {Heger}, A., {Gladysz}, W., {et~al.} 2016, \aap, 594, A97

\bibitem[{{Belczynski} {et~al.}(2002){Belczynski}, {Kalogera}, \&
  {Bulik}}]{2002ApJ...572..407B}
{Belczynski}, K., {Kalogera}, V., \& {Bulik}, T. 2002, \apj, 572, 407

\bibitem[{{Belczynski} {et~al.}(2008){Belczynski}, {Kalogera}, {Rasio}, {Taam},
  {Zezas}, {Bulik}, {Maccarone}, \& {Ivanova}}]{2008ApJS..174..223B}
{Belczynski}, K., {Kalogera}, V., {Rasio}, F.~A., {et~al.} 2008, \apjs, 174,
  223

\bibitem[{Belczynski {et~al.}(2017)Belczynski, Klencki, Fields, Olejak, Berti,
  Meynet, Fryer, Holz, O'Shaughnessy, Brown, Bulik, Leung, Nomoto, Madau,
  Hirschi, Jones, Mondal, Chruslinska, Drozda, Gerosa, Doctor, Giersz, Ekstrom,
  Georgy, Askar, Wysocki, Natan, Farr, Wiktorowicz, Miller, Farr, \&
  Lasota}]{belczynski2017evolutionary}
Belczynski, K., Klencki, J., Fields, C.~E., {et~al.} 2017, The evolutionary
  roads leading to low effective spins, high black hole masses, and O1/O2 rates
  of LIGO/Virgo binary black holes

\bibitem[{{Belczynski} {et~al.}(2017){Belczynski}, {Klencki}, {Meynet},
  {Fryer}, {Brown}, {Chruslinska}, {Gladysz}, {O'Shaughnessy}, {Bulik}, \&
  {Berti}}]{2017arXiv170607053B}
{Belczynski}, K., {Klencki}, J., {Meynet}, G., {et~al.} 2017, arXiv e-prints,
  arXiv:1706.07053

\bibitem[{{Bensby} {et~al.}(2018){Bensby}, {Feltzing}, {Gould}, {Yee},
  {Johnson}, {Asplund}, {Mel{\'e}ndez}, {Lucatello}, \&
  {Howes}}]{2018IAUS..334...86B}
{Bensby}, T., {Feltzing}, S., {Gould}, A., {et~al.} 2018, in IAU Symposium,
  Vol. 334, Rediscovering Our Galaxy, ed. C.~{Chiappini}, I.~{Minchev},
  E.~{Starkenburg}, \& M.~{Valentini}, 86--89

\bibitem[{{Bird} {et~al.}(2016){Bird}, {Cholis}, {Mu{\~n}oz},
  {Ali-Ha{\"\i}moud}, {Kamionkowski}, {Kovetz}, {Raccanelli}, \&
  {Riess}}]{2016PhRvL.116t1301B}
{Bird}, S., {Cholis}, I., {Mu{\~n}oz}, J.~B., {et~al.} 2016, \prl, 116, 201301

\bibitem[{{Blaauw}(1961)}]{1961BAN....15..265B}
{Blaauw}, A. 1961, \bain, 15, 265

\bibitem[{{Brown} \& {Bethe}(1994)}]{1994ApJ...423..659B}
{Brown}, G.~E. \& {Bethe}, H.~A. 1994, \apj, 423, 659

\bibitem[{{Bullock} \& {Johnston}(2005)}]{2005ApJ...635..931B}
{Bullock}, J.~S. \& {Johnston}, K.~V. 2005, \apj, 635, 931

\bibitem[{{Casagrande} {et~al.}(2011){Casagrande}, {Sch{\"o}nrich}, {Asplund},
  {Cassisi}, {Ram{\'{\i}}rez}, {Mel{\'e}ndez}, {Bensby}, \&
  {Feltzing}}]{2011A&A...530A.138C}
{Casagrande}, L., {Sch{\"o}nrich}, R., {Asplund}, M., {et~al.} 2011, \aap, 530,
  A138

\bibitem[{{Casares}(2007)}]{2007IAUS..238....3C}
{Casares}, J. 2007, in IAU Symposium, Vol. 238, Black Holes from Stars to
  Galaxies -- Across the Range of Masses, ed. V.~{Karas} \& G.~{Matt}, 3--12

\bibitem[{{Casares} \& {Jonker}(2014)}]{2014SSRv..183..223C}
{Casares}, J. \& {Jonker}, P.~G. 2014, \ssr, 183, 223

\bibitem[{{Chiba} \& {Beers}(2000)}]{2000AJ....119.2843C}
{Chiba}, M. \& {Beers}, T.~C. 2000, \aj, 119, 2843

\bibitem[{{Cignoni} {et~al.}(2006){Cignoni}, {Degl'Innocenti}, {Prada Moroni},
  \& {Shore}}]{2006A&A...459..783C}
{Cignoni}, M., {Degl'Innocenti}, S., {Prada Moroni}, P.~G., \& {Shore}, S.~N.
  2006, \aap, 459, 783

\bibitem[{{de Mink} \& {Belczynski}(2015)}]{2015ApJ...814...58D}
{de Mink}, S.~E. \& {Belczynski}, K. 2015, \apj, 814, 58

\bibitem[{{Demorest} {et~al.}(2010){Demorest}, {Pennucci}, {Ransom}, {Roberts},
  \& {Hessels}}]{2010Natur.467.1081D}
{Demorest}, P.~B., {Pennucci}, T., {Ransom}, S.~M., {Roberts}, M.~S.~E., \&
  {Hessels}, J.~W.~T. 2010, \nat, 467, 1081

\bibitem[{{Dominik} {et~al.}(2013){Dominik}, {Belczynski}, {Fryer}, {Holz},
  {Berti}, {Bulik}, {Mand el}, \& {O'Shaughnessy}}]{2013ApJ...779...72D}
{Dominik}, M., {Belczynski}, K., {Fryer}, C., {et~al.} 2013, \apj, 779, 72

\bibitem[{{Dominik} {et~al.}(2012){Dominik}, {Belczynski}, {Fryer}, {Holz},
  {Berti}, {Bulik}, {Mandel}, \& {O'Shaughnessy}}]{2012ApJ...759...52D}
{Dominik}, M., {Belczynski}, K., {Fryer}, C., {et~al.} 2012, \apj, 759, 52

\bibitem[{{Duch{\^e}ne} \& {Kraus}(2013)}]{2013ARA&A..51..269D}
{Duch{\^e}ne}, G. \& {Kraus}, A. 2013, \araa, 51, 269

\bibitem[{{Dvorkin} {et~al.}(2016){Dvorkin}, {Vangioni}, {Silk}, {Uzan}, \&
  {Olive}}]{2016MNRAS.461.3877D}
{Dvorkin}, I., {Vangioni}, E., {Silk}, J., {Uzan}, J.-P., \& {Olive}, K.~A.
  2016, \mnras, 461, 3877

\bibitem[{Eggleton \& Kiseleva-Eggleton(2001)}]{Eggleton_2001}
Eggleton, P.~P. \& Kiseleva-Eggleton, L. 2001, The Astrophysical Journal, 562,
  1012

\bibitem[{{Fryer} {et~al.}(2012{\natexlab{a}}){Fryer}, {Belczynski},
  {Wiktorowicz}, {Dominik}, {Kalogera}, \& {Holz}}]{Fryer1204}
{Fryer}, C.~L., {Belczynski}, K., {Wiktorowicz}, G., {et~al.}
  2012{\natexlab{a}}, \apj, 749, 91

\bibitem[{{Fryer} {et~al.}(2012{\natexlab{b}}){Fryer}, {Belczynski},
  {Wiktorowicz}, {Dominik}, {Kalogera}, \& {Holz}}]{2012ApJ...749...91F}
{Fryer}, C.~L., {Belczynski}, K., {Wiktorowicz}, G., {et~al.}
  2012{\natexlab{b}}, \apj, 749, 91

\bibitem[{{Gao} {et~al.}(2014){Gao}, {Liu}, {Zhang}, {Justham}, {Deng}, \&
  {Yang}}]{2014ApJ...788L..37G}
{Gao}, S., {Liu}, C., {Zhang}, X., {et~al.} 2014, \apjl, 788, L37

\bibitem[{{Gerhard}(2001)}]{2001ASPC..230...21G}
{Gerhard}, O.~E. 2001, in Astronomical Society of the Pacific Conference
  Series, Vol. 230, Galaxy Disks and Disk Galaxies, ed. J.~G. {Funes} \& E.~M.
  {Corsini}, 21--30

\bibitem[{Glebbeek {et~al.}(2013)Glebbeek, Gaburov, Portegies~Zwart, \&
  Pols}]{Glebbeek_2013}
Glebbeek, E., Gaburov, E., Portegies~Zwart, S., \& Pols, O.~R. 2013, Monthly
  Notices of the Royal Astronomical Society, 434, 3497–3510

\bibitem[{{Grand} {et~al.}(2019){Grand}, {Deason}, {White}, {Simpson},
  {G{\'o}mez}, {Marinacci}, \& {Pakmor}}]{2019MNRAS.487L..72G}
{Grand}, R. J.~J., {Deason}, A.~J., {White}, S. D.~M., {et~al.} 2019, \mnras,
  487, L72

\bibitem[{{Haywood} {et~al.}(2015){Haywood}, {Di Matteo}, \&
  {Lehnert}}]{2015cdem.confE...3H}
{Haywood}, M., {Di Matteo}, P., \& {Lehnert}, M. 2015, in Chemical and
  dynamical evolution of the Milky Way and Local Group, 3

\bibitem[{{Haywood} {et~al.}(2013){Haywood}, {Di Matteo}, {Lehnert}, {Katz}, \&
  {G{\'o}mez}}]{2013A&A...560A.109H}
{Haywood}, M., {Di Matteo}, P., {Lehnert}, M.~D., {Katz}, D., \& {G{\'o}mez},
  A. 2013, \aap, 560, A109

\bibitem[{{Hobbs} {et~al.}(2006){Hobbs}, {Lorimer}, {Lyne}, \&
  {Kramer}}]{2006yCat..73600974H}
{Hobbs}, G., {Lorimer}, D.~R., {Lyne}, A.~G., \& {Kramer}, M. 2006, VizieR
  Online Data Catalog, J/MNRAS/360/974

\bibitem[{{Hurley} {et~al.}(2000){Hurley}, {Pols}, \&
  {Tout}}]{2000MNRAS.315..543H}
{Hurley}, J.~R., {Pols}, O.~R., \& {Tout}, C.~A. 2000, \mnras, 315, 543

\bibitem[{Hurley {et~al.}(2002)Hurley, Tout, \& Pols}]{Hurley_2002}
Hurley, J.~R., Tout, C.~A., \& Pols, O.~R. 2002, Monthly Notices of the Royal
  Astronomical Society, 329, 897–928

\bibitem[{{Igoshev} \& {Perets}(2019)}]{2019MNRAS.486.4098I}
{Igoshev}, A.~P. \& {Perets}, H.~B. 2019, \mnras, 486, 4098

\bibitem[{{Ivanova} \& {Taam}(2004)}]{Ivanova0402}
{Ivanova}, N. \& {Taam}, R.~E. 2004, \apj, 601, 1058

\bibitem[{J.~C.~Lombardi {et~al.}(2002)J.~C.~Lombardi, Warren, Rasio, Sills, \&
  Warren}]{Lombardi_Jr__2002}
J.~C.~Lombardi, J., Warren, J.~S., Rasio, F.~A., Sills, A., \& Warren, A.~R.
  2002, The Astrophysical Journal, 568, 939

\bibitem[{{Janka}(2017)}]{2017hsn..book.1095J}
{Janka}, H.-T. 2017, {Neutrino-Driven Explosions}, 1095

\bibitem[{{Kalogera}(1996)}]{1996ApJ...471..352K}
{Kalogera}, V. 1996, \apj, 471, 352

\bibitem[{{Kalogera} \& {Baym}(1996)}]{1996ApJ...470L..61K}
{Kalogera}, V. \& {Baym}, G. 1996, \apjl, 470, L61

\bibitem[{{Khokhlov} {et~al.}(2018){Khokhlov}, {Miroshnichenko}, {Zharikov},
  {Manset}, {Arkharov}, {Efimova}, {Klimanov}, {Larionov}, {Kusakin},
  {Kokumbaeva}, {Omarov}, {Kuratov}, {Kuratova}, {Rudy}, {Laag}, {Crawford},
  {Swift}, {Puetter}, {Perry}, {Chojnowski}, {Agishev}, {Caton}, {Hawkins},
  {Smith}, {Reichart}, {Kouprianov}, \& {Haislip}}]{2018ApJ...856..158K}
{Khokhlov}, S.~A., {Miroshnichenko}, A.~S., {Zharikov}, S.~V., {et~al.} 2018,
  \apj, 856, 158

\bibitem[{{Klencki} {et~al.}(2018){Klencki}, {Moe}, {Gladysz}, {Chruslinska},
  {Holz}, \& {Belczynski}}]{2018A&A...619A..77K}
{Klencki}, J., {Moe}, M., {Gladysz}, W., {et~al.} 2018, \aap, 619, A77

\bibitem[{{Klencki} {et~al.}(2017){Klencki}, {Wiktorowicz}, {G{\l}adysz}, \&
  {Belczynski}}]{2017MNRAS.469.3088K}
{Klencki}, J., {Wiktorowicz}, G., {G{\l}adysz}, W., \& {Belczynski}, K. 2017,
  \mnras, 469, 3088

\bibitem[{{Kobayashi} \& {Nakasato}(2011)}]{2011ApJ...729...16K}
{Kobayashi}, C. \& {Nakasato}, N. 2011, \apj, 729, 16

\bibitem[{{Kobulnicky} {et~al.}(2014){Kobulnicky}, {Kiminki}, {Lundquist},
  {Burke}, {Chapman}, {Keller}, {Lester}, {Rolen}, {Topel}, {Bhattacharjee},
  {Smullen}, {Vargas {\'A}lvarez}, {Runnoe}, {Dale}, \&
  {Brotherton}}]{2014ApJS..213...34K}
{Kobulnicky}, H.~A., {Kiminki}, D.~C., {Lundquist}, M.~J., {et~al.} 2014,
  \apjs, 213, 34

\bibitem[{{Kochanek} {et~al.}(2014){Kochanek}, {Adams}, \&
  {Belczynski}}]{2014MNRAS.443.1319K}
{Kochanek}, C.~S., {Adams}, S.~M., \& {Belczynski}, K. 2014, \mnras, 443, 1319

\bibitem[{Korol {et~al.}(2018)Korol, Rossi, \&
  Barausse}]{korol2018constraining}
Korol, V., Rossi, E.~M., \& Barausse, E. 2018, Constraining the Milky Way
  potential with Double White Dwarfs

\bibitem[{{Kravtsov} \& {Gnedin}(2005)}]{2005ApJ...623..650K}
{Kravtsov}, A.~V. \& {Gnedin}, O.~Y. 2005, \apj, 623, 650

\bibitem[{{Kroupa}(2002)}]{2002Sci...295...82K}
{Kroupa}, P. 2002, Science, 295, 82

\bibitem[{{Kroupa} {et~al.}(1993){Kroupa}, {Tout}, \&
  {Gilmore}}]{1993MNRAS.262..545K}
{Kroupa}, P., {Tout}, C.~A., \& {Gilmore}, G. 1993, \mnras, 262, 545

\bibitem[{{Kruijssen} \& {Mieske}(2009)}]{2009A&A...500..785K}
{Kruijssen}, J.~M.~D. \& {Mieske}, S. 2009, \aap, 500, 785

\bibitem[{{Lamberts} {et~al.}(2018){Lamberts}, {Garrison-Kimmel}, {Hopkins},
  {Quataert}, {Bullock}, {Faucher-Gigu{\`e}re}, {Wetzel}, {Kere{\v s}},
  {Drango}, \& {Sanderson}}]{2018MNRAS.480.2704L}
{Lamberts}, A., {Garrison-Kimmel}, S., {Hopkins}, P.~F., {et~al.} 2018, \mnras,
  480, 2704

\bibitem[{{Leung} {et~al.}(2019){Leung}, {Nomoto}, \&
  {Blinnikov}}]{2019arXiv190111136L}
{Leung}, S.-C., {Nomoto}, K., \& {Blinnikov}, S. 2019, arXiv e-prints,
  arXiv:1901.11136

\bibitem[{Li(2016)}]{li2016modelling}
Li, E. 2016, Modelling mass distribution of the Milky Way galaxy using Gaia
  billion-star map

\bibitem[{{Licquia} \& {Newman}(2015)}]{Licquia1506}
{Licquia}, T.~C. \& {Newman}, J.~A. 2015, \apj, 806, 96

\bibitem[{{Linares} {et~al.}(2018){Linares}, {Shahbaz}, \&
  {Casares}}]{2018cosp...42E2021L}
{Linares}, M., {Shahbaz}, T., \& {Casares}, J. 2018, in COSPAR Meeting,
  Vol.~42, 42nd COSPAR Scientific Assembly, E1.10--9--18

\bibitem[{Liu {et~al.}(2019)Liu, Zhang, Howard, Bai, Lu, Soria, Justham, Li,
  Zheng, Wang, \& et~al.}]{Liu_2019}
Liu, J., Zhang, H., Howard, A.~W., {et~al.} 2019, Nature, 575, 618–621

\bibitem[{{Liu} {et~al.}(2018){Liu}, {Du}, {Newberg}, {Chen}, {Wu}, {Ma},
  {Zhou}, {Cao}, {Hou}, {Wang}, \& {Zhang}}]{2018ApJ...862..163L}
{Liu}, S., {Du}, C., {Newberg}, H.~J., {et~al.} 2018, \apj, 862, 163

\bibitem[{{Lombardi} {et~al.}(2006){Lombardi}, {Proulx}, {Dooley}, {Theriault},
  {Ivanova}, \& {Rasio}}]{2006ApJ...640..441L}
{Lombardi}, J.~C., J., {Proulx}, Z.~F., {Dooley}, K.~L., {et~al.} 2006, \apj,
  640, 441

\bibitem[{{Mennekens} \& {Vanbeveren}(2014)}]{2014A&A...564A.134M}
{Mennekens}, N. \& {Vanbeveren}, D. 2014, \aap, 564, A134

\bibitem[{{Monari} {et~al.}(2018){Monari}, {Famaey}, {Carrillo}, {Piffl},
  {Steinmetz}, {Wyse}, {Anders}, {Chiappini}, \&
  {Jan{\ss}en}}]{2018A&A...616L...9M}
{Monari}, G., {Famaey}, B., {Carrillo}, I., {et~al.} 2018, \aap, 616, L9

\bibitem[{{Monroy-Rodr{\'\i}guez} \& {Allen}(2014)}]{2014ApJ...790..159M}
{Monroy-Rodr{\'\i}guez}, M.~A. \& {Allen}, C. 2014, \apj, 790, 159

\bibitem[{{Morrison} {et~al.}(2000){Morrison}, {Mateo}, {Olszewski}, {Harding},
  {Dohm-Palmer}, {Freeman}, {Norris}, \& {Morita}}]{2000AJ....119.2254M}
{Morrison}, H.~L., {Mateo}, M., {Olszewski}, E.~W., {et~al.} 2000, \aj, 119,
  2254

\bibitem[{{Raghavan} {et~al.}(2010){Raghavan}, {McAlister}, {Henry}, {Latham},
  {Marcy}, {Mason}, {Gies}, {White}, \& {ten Brummelaar}}]{2010ApJS..190....1R}
{Raghavan}, D., {McAlister}, H.~A., {Henry}, T.~J., {et~al.} 2010, \apjs, 190,
  1

\bibitem[{Rix \& Bovy(2013)}]{Rix_2013}
Rix, H.-W. \& Bovy, J. 2013, The Astronomy and Astrophysics Review, 21

\bibitem[{{Sana} {et~al.}(2012){Sana}, {de Mink}, {de Koter}, {Langer},
  {Evans}, {Gieles}, {Gosset}, {Izzard}, {Le Bouquin}, \&
  {Schneider}}]{Sana1207}
{Sana}, H., {de Mink}, S.~E., {de Koter}, A., {et~al.} 2012, Science, 337, 444

\bibitem[{{Sana} {et~al.}(2014){Sana}, {Le Bouquin}, {Lacour}, {Berger},
  {Duvert}, {Gauchet}, {Norris}, {Olofsson}, {Pickel}, {Zins}, {Absil}, {de
  Koter}, {Kratter}, {Schnurr}, \& {Zinnecker}}]{2014ApJS..215...15S}
{Sana}, H., {Le Bouquin}, J.~B., {Lacour}, S., {et~al.} 2014, \apjs, 215, 15

\bibitem[{{Shapiro} \& {Teukolsky}(1983)}]{1983bhwd.book.....S}
{Shapiro}, S.~L. \& {Teukolsky}, S.~A. 1983, {Black holes, white dwarfs, and
  neutron stars: The physics of compact objects}

\bibitem[{{Siegel} {et~al.}(2002){Siegel}, {Majewski}, {Reid}, \&
  {Thompson}}]{2002ApJ...578..151S}
{Siegel}, M.~H., {Majewski}, S.~R., {Reid}, I.~N., \& {Thompson}, I.~B. 2002,
  \apj, 578, 151

\bibitem[{{Soubiran} \& {Girard}(2005)}]{2005A&A...438..139S}
{Soubiran}, C. \& {Girard}, P. 2005, \aap, 438, 139

\bibitem[{{Spera} {et~al.}(2015){Spera}, {Mapelli}, \&
  {Bressan}}]{2015MNRAS.451.4086S}
{Spera}, M., {Mapelli}, M., \& {Bressan}, A. 2015, \mnras, 451, 4086

\bibitem[{Swihart {et~al.}(2017)Swihart, Strader, Johnson, Cheung, Sand,
  Chomiuk, Wasserman, Larsen, Brodie, Simonian, Tremou, Shishkovsky, Reichart,
  \& Haislip}]{Swihart_2017}
Swihart, S.~J., Strader, J., Johnson, T.~J., {et~al.} 2017, The Astrophysical
  Journal, 851, 31

\bibitem[{{Tanikawa} {et~al.}(2019){Tanikawa}, {Kinugawa}, {Kumamoto}, \&
  {Fujii}}]{2019arXiv191204509T}
{Tanikawa}, A., {Kinugawa}, T., {Kumamoto}, J., \& {Fujii}, M.~S. 2019, arXiv
  e-prints, arXiv:1912.04509

\bibitem[{{The LIGO Scientific Collaboration} {et~al.}(2018){The LIGO
  Scientific Collaboration}, {the Virgo Collaboration}, {Abbott}, {Abbott},
  {Abbott}, {Abraham}, {Acernese}, {Ackley}, {Adams}, \&
  {Adhikari}}]{2018arXiv181112907T}
{The LIGO Scientific Collaboration}, {the Virgo Collaboration}, {Abbott},
  B.~P., {et~al.} 2018, arXiv e-prints, arXiv:1811.12907

\bibitem[{{Thompson} {et~al.}(2019){Thompson}, {Kochanek}, {Stanek}, {Badenes},
  {Post}, {Jayasinghe}, {Latham}, {Bieryla}, {Esquerdo}, {Berlind}, {Calkins},
  {Tayar}, {Lindegren}, {Johnson}, {Holoien}, {Auchettl}, \&
  {Covey}}]{2019Sci...366..637T}
{Thompson}, T.~A., {Kochanek}, C.~S., {Stanek}, K.~Z., {et~al.} 2019, Science,
  366, 637

\bibitem[{{Timmes} {et~al.}(1996){Timmes}, {Woosley}, \&
  {Weaver}}]{1996ApJ...457..834T}
{Timmes}, F.~X., {Woosley}, S.~E., \& {Weaver}, T.~A. 1996, \apj, 457, 834

\bibitem[{{Tisserand} {et~al.}(2007){Tisserand}, {Le Guillou}, {Afonso},
  {Albert}, {Andersen}, {Ansari}, {Aubourg}, {Bareyre}, {Beaulieu}, {Charlot},
  {Coutures}, {Ferlet}, {Fouqu{\'e}}, {Glicenstein}, {Goldman}, {Gould},
  {Graff}, {Gros}, {Haissinski}, {Hamadache}, {de Kat}, {Lasserre}, {Lesquoy},
  {Loup}, {Magneville}, {Marquette}, {Maurice}, {Maury}, {Milsztajn}, {Moniez},
  {Palanque-Delabrouille}, {Perdereau}, {Rahal}, {Rich}, {Spiro},
  {Vidal-Madjar}, {Vigroux}, {Zylberajch}, \& {EROS-2
  Collaboration}}]{2007A&A...469..387T}
{Tisserand}, P., {Le Guillou}, L., {Afonso}, C., {et~al.} 2007, \aap, 469, 387

\bibitem[{{Tsuna} {et~al.}(2018){Tsuna}, {Kawanaka}, \&
  {Totani}}]{2018MNRAS.477..791T}
{Tsuna}, D., {Kawanaka}, N., \& {Totani}, T. 2018, \mnras, 477, 791

\bibitem[{{Tutukov} {et~al.}(1992){Tutukov}, {Yungelson}, \&
  {Iben}}]{1992ApJ...386..197T}
{Tutukov}, A.~V., {Yungelson}, L.~R., \& {Iben}, Jr., I. 1992, \apj, 386, 197

\bibitem[{{Tylenda} \& {Kami{\'n}ski}(2016)}]{2016A&A...592A.134T}
{Tylenda}, R. \& {Kami{\'n}ski}, T. 2016, \aap, 592, A134

\bibitem[{{van den Heuvel}(1992)}]{1992eocm.rept...29V}
{van den Heuvel}, E.~P.~J. 1992, {Endpoints of stellar evolution: The incidence
  of stellar mass black holes in the galaxy}, Tech. rep.

\bibitem[{Vanbeveren \& Donder(2010)}]{VANBEVEREN201050}
Vanbeveren, D. \& Donder, E.~D. 2010, New Astronomy Reviews, 54, 50 ,
  proceedings: A Life With Stars

\bibitem[{Voss \& Tauris(2003)}]{10.1046/j.1365-8711.2003.06616.x}
Voss, R. \& Tauris, T.~M. 2003, Monthly Notices of the Royal Astronomical
  Society, 342, 1169

\bibitem[{{Wang} {et~al.}(2015){Wang}, {Han}, {Cooper}, {Cole}, {Frenk}, \&
  {Lowing}}]{2015MNRAS.453..377W}
{Wang}, W., {Han}, J., {Cooper}, A.~P., {et~al.} 2015, \mnras, 453, 377

\bibitem[{{Webbink}(1984)}]{1984ApJ...277..355W}
{Webbink}, R.~F. 1984, \apj, 277, 355

\bibitem[{{Wiktorowicz} {et~al.}(2019){Wiktorowicz}, {Wyrzykowski},
  {Chruslinska}, {Klencki}, {Rybicki}, \& {Belczynski}}]{2019arXiv190711431W}
{Wiktorowicz}, G., {Wyrzykowski}, {\L}., {Chruslinska}, M., {et~al.} 2019,
  arXiv e-prints, arXiv:1907.11431

\bibitem[{{Woosley}(2017)}]{2017ApJ...836..244W}
{Woosley}, S.~E. 2017, \apj, 836, 244

\bibitem[{{Wyrzykowski} {et~al.}(2016{\natexlab{a}}){Wyrzykowski},
  {Kostrzewa-Rutkowska}, \& {Rybicki}}]{2016pas..conf..121W}
{Wyrzykowski}, {\L}., {Kostrzewa-Rutkowska}, Z., \& {Rybicki}, K.
  2016{\natexlab{a}}, in 37th Meeting of the Polish Astronomical Society, ed.
  A.~{R{\'o}{\.z}a{\'n}ska} \& M.~{Bejger}, Vol.~3, 121--124

\bibitem[{{Wyrzykowski} {et~al.}(2016{\natexlab{b}}){Wyrzykowski},
  {Kostrzewa-Rutkowska}, {Skowron}, {Rybicki}, {Mr{\'o}z}, {Koz{\l}owski},
  {Udalski}, {Szyma{\'n}ski}, {Pietrzy{\'n}ski}, {Soszy{\'n}ski}, {Ulaczyk},
  {Pietrukowicz}, {Poleski}, {Pawlak}, {I{\l}kiewicz}, \&
  {Rattenbury}}]{2016MNRAS.458.3012W}
{Wyrzykowski}, {\L}., {Kostrzewa-Rutkowska}, Z., {Skowron}, J., {et~al.}
  2016{\natexlab{b}}, \mnras, 458, 3012

\bibitem[{{Wyrzykowski} \& {Mandel}(2019)}]{2019arXiv190407789W}
{Wyrzykowski}, {\L}. \& {Mandel}, I. 2019, arXiv e-prints, arXiv:1904.07789

\bibitem[{{Wyrzykowski} {et~al.}(2011){Wyrzykowski}, {Skowron}, {Koz{\l}owski},
  {Udalski}, {Szyma{\'n}ski}, {Kubiak}, {Pietrzy{\'n}ski}, {Soszy{\'n}ski},
  {Szewczyk}, {Ulaczyk}, {Poleski}, \& {Tisserand}}]{2011MNRAS.416.2949W}
{Wyrzykowski}, L., {Skowron}, J., {Koz{\l}owski}, S., {et~al.} 2011, \mnras,
  416, 2949

\bibitem[{{Yanny} {et~al.}(2000){Yanny}, {Newberg}, {Kent},
  {Laurent-Muehleisen}, {Pier}, {Richards}, {Stoughton}, {Anderson}, {Annis},
  {Brinkmann}, {Chen}, {Csabai}, {Doi}, {Fukugita}, {Hennessy}, {Ivezi{\'c}},
  {Knapp}, {Lupton}, {Munn}, {Nash}, {Rockosi}, {Schneider}, {Smith}, \&
  {York}}]{2000ApJ...540..825Y}
{Yanny}, B., {Newberg}, H.~J., {Kent}, S., {et~al.} 2000, \apj, 540, 825

\bibitem[{{Yoo} {et~al.}(2004){Yoo}, {Chanam{\'e}}, \&
  {Gould}}]{2004ApJ...601..311Y}
{Yoo}, J., {Chanam{\'e}}, J., \& {Gould}, A. 2004, \apj, 601, 311

\end{thebibliography}

\end{document}